\documentclass[aps,prd,10pt,amsmath,amssymb,nofootinbib,notitlepage,superscriptaddress,twocolumn]{revtex4-2}

\usepackage[utf8]{inputenc}
\usepackage{comment}
\usepackage{amsmath}
\usepackage{braket}
\usepackage{mathrsfs}
\usepackage{mathtools}
\usepackage{array}
\usepackage{fancyhdr}
\usepackage{bm}
\usepackage{multirow}
\usepackage{makecell}

\usepackage{ragged2e}

\usepackage{diagbox}
\usepackage{nicematrix}
\usepackage{tikz}
\usetikzlibrary{patterns}
\usetikzlibrary{arrows.meta}

\usepackage[justification=justified]{subcaption}

\allowdisplaybreaks

\definecolor{RedWine}{rgb}{0.743,0,0}
\definecolor{RoyalBlue}{rgb}{0.25,.41,.88}
\definecolor{ForestGreen}{rgb}{.13,.54,.13}

\newcommand{\aap}{A\&A}
\newcommand{\apjl}{Astrophys. J. Lett.}

\usepackage{tensor}
\usepackage{xcolor}
\usepackage{slashed}
\usepackage{dsfont} 
\usepackage{bbold}

\usepackage[backref,breaklinks,colorlinks,citecolor=ForestGreen]{hyperref}

\begin{document}

\title{\texorpdfstring{Plasma flow in force-free magnetospheres: \\ two-fluid model near pulsars and black holes}{Plasma flow in force-free magnetospheres: two-fluid model near pulsars and black holes}}

\author{Morifumi Mizuno}
\affiliation{
Department of Physics, University of Arizona, Tucson, Arizona 85721, USA
}

\author{Samuel E. Gralla}
\affiliation{
Department of Physics, University of Arizona, Tucson, Arizona 85721, USA
}

\author{Alexander Philippov}
\affiliation{
Department of Physics, University of Maryland, College Park, 20742 Maryland, USA
}
\affiliation{
Institute for Research in Electronics and Applied Physics, University of Maryland,
College Park, Maryland 20742, USA
}

\begin{abstract}
Force-free electrodynamics describes the electromagnetic field of the magnetically dominated plasma found near pulsars and active black holes, but gives no information about the underlying particles that ultimately produce the observable emission. Working in the two-fluid approximation, we show how particles can be ``painted on'' to a force-free solution as a function of boundary conditions that encode the particle output of ``gap regions'' where the force-free approximation does not hold. These boundary conditions also determine the leading parallel electric field in the entire magnetosphere.  Our treatment holds in a general (possibly curved) spacetime and is phrased in language intrinsic to the 1+1 dimensional ``field sheet spacetimes'' experienced by particles stuck to magnetic field lines.  Besides the new results, this provides an elegant formulation of some standard equations; for example, we show that the zero-gyroradius guiding center approximation is just the Lorentz force law on the field sheet.  We derive a general perturbative method and apply it to pulsar and black hole magnetospheres with radial magnetic fields to produce fully analytic models that capture key features of the full problem. When applied to more realistic magnetic field configurations together with simulation-informed boundary conditions for the gap regions, this approach has the potential to provide global magnetosphere models without the need for global particle-in-cell simulations.
\end{abstract}
\maketitle

\section{Introduction}

Force-free electrodynamics (FFE) \citep{Uchida1997aug,Uchida1997oct,Uchida1998jun,Komissarov2002nov,Gralla.Jacobson2014dec} is a universal description of magnetically-dominated plasma physics that follows purely from conservation laws---see appendix \ref{sec:FFE} for a brief summary. One assumes that there is sufficient charged matter to affect the electromagnetic field, but that the matter itself is energetically negligible. One then obtains a deterministic theory of the electromagnetic field alone,
\begin{align}\begin{gathered}
    \nabla_a F^{ab} F_{bc} = 0, \qquad \nabla_a \tilde{F}^{ab} = 0, \\ F_{ab}\tilde{F}^{ab} = 0, \qquad  F_{ab} F^{ab} > 0. \end{gathered}\label{FFE}
\end{align}
Here $F_{ab}$ is the Maxwell field propagating in a (possibly curved) spacetime metric $g_{ab}$ with compatible derivative operator $\nabla_a$ and $\tilde{F}_{ab}=\frac{1}{2}\epsilon_{abcd}F^{cd}$ is the dual field strength tensor. The first equation is the vanishing of the Lorentz force density $F_{ab}j^b=0$ (recalling the Maxwell equation $j^a=\nabla_b F^{ab}$), and gives the theory its name. The second equation is conservation of magnetic flux. The third equation is the degeneracy of the electromagnetic field ($\vec{E}\cdot\vec{B}=0$), which is interpreted as complete and instantaneous screening of the electric fields by a conducting plasma. The last equation is magnetic domination $B^2>E^2$, which is required for the evolution to be deterministic \citep{Palenzuela.Bona.ea2011jun,Pfeiffer.MacFadyen2013jul,Carrasco.Reula2016apr}.

The striking feature of FFE is the complete absence of plasma degrees of freedom.  The presence of charged matter is essential to the validity of these equations and key to their physical interpretation, but the detailed description of the matter is unnecessary---in fact, \textit{irrelevant}---for determining the electromagnetic field. 
  The force-free equations are totally insensitive to the underlying nature of the plasma: they emerge as strong-field limits of both traditional \cite{Goldreich.Julian1969aug,Scharlemann1974oct} and  more exotic \citep{Gralla2019may,Gralla.Iqbal2019may,2022ScPP...12...78I,2022ScPP...12...86I} descriptions, and can determine the electromagnetic field even when the underlying plasma description is unknown. 

This universality becomes a limitation, however, when one needs to know the behavior of the plasma.  A force-free solution by itself offers no clue as to what the plasma is doing, and one must turn to the specific theory of plasma most relevant to the physics.  Even though universality is now lost, the force-free limit still offers some simplification, since the electromagnetic field may be determined with FFE \textit{before} the plasma behavior is determined.  Thus, for any given description of plasma, one expects some kind of staggered perturbation theory in the force-free limit, wherein plasma behavior is ``painted'' on to the force-free solution after it is determined.  In addition to the practical benefits of such a simplification, the perturbative approach can help delineate the regime of validity of the force-free approximation and give insight into the nature of the first corrections. 

One of the most important applications of FFE is the description of pulsar \cite{Gold1968may} and black hole \cite{Blandford.Znajek1977jul} magnetospheres in the regime where pair production is plentiful.  The basic physics is that the rotation of the compact object in a magnetic field produces huge voltages that trigger particle acceleration and pair production (via various QED processes) until there are sufficient pairs to screen the accelerating electric field.  Such screening occurs at an energetically negligible charge density, putting the resulting plasma squarely in the force-free regime.  
 However, the charges are born at high velocity and ultimately escape the magnetosphere, meaning that ``gap regions'' of unscreened electric field will open up (or exist continuously) and replenish the plasma.  
 
 Numerical solutions of FFE are by now routine \citep[e.g.,][]{Spitkovsky2006aug,Kalapotharakos.Contopoulos2009mar,Petri2012jul}, allowing the large-scale electromagnetic field to be determined for a given set of magnetic field boundary conditions.\footnote{Note that some non-force-free prescription is required to handle current sheet regions; options include manually limiting the electric field to be weaker than the magnetic field \cite{Spitkovsky2006aug} or introducing resistivity into the Ohm's law \citep[e.g.,][]{Palenzuela:2010xn,Li2012}. More consistent treatment in the fluid limit requires including plasma pressure, as is done in the magnetohydrodynamic approximation \citep[e.g.,][]{Tchekhovskoy.Spitkovsky.ea2013aug}}  However, much less is understood about the production and dynamics of the underlying particles.  Although there has been tremendous progress using particle-in-cell (PIC) simulations in recent years (see \cite{{Philippov.Kramer2022}} for a review of recent work), the high computational cost still limits the dynamic range of global simulations and forces a focus on just a few canonical magnetic field configurations.  

 However, in the regime of plentiful pair production, a global PIC simulation would seem to be ``overkill'' since the vast majority of the magnetosphere is force-free.  Instead, it should be possible to determine the global electromagnetic field with FFE and then paint on the plasma according to assumptions about the gap regions.  These assumptions could be derived from local high-resolution PIC simulations and/or justified by comparison with a handful of global PIC simulations.  
 
 As a step toward this kind of staggered magnetosphere modeling, in this paper we study the force-free limit of two-fluid plasma theory.  We derive a general method for painting particles on an arbitrary force-free solution in an arbitrary curved spacetime, show how the equations can be integrated analytically if there is a symmetry adapted to field line motion (a case that includes inclined rotating magnetospheres).  We illustrate the method in aligned pulsar and black hole magnetospheres with radial field lines.

Our approach generalizes prior work of Sharlemann \cite{Scharlemann1974oct} and Petrova \cite{Petrova2015jan,Petrova2017may} on aligned pulsar magnetospheres in flat spacetime.  As in these references, we take the zero-gyroradius limit of two-fluid theory via a formal expansion in the mass $m$ of the charge-carriers.  We find that the leading-order field is described by FFE, while the leading-order particles move along magnetic field lines in the presence of a small parallel electric field.  The requirement that both species feel the same electric field allows the equations to be decoupled, so that both the particle motion and the electric field may be self-consistently determined from boundary conditions that encode the gap physics. 

Our treatment is geometrical in nature, making important use of the 1+1 dimensional magnetic ``field sheets'' (spacetime manifolds of magnetic field lines) defined by the force-free background.  This approach simplifies and clarifies the physics.  For example, we find that particles follow the Lorentz force law on the field sheets, providing an elegant reformulation of the traditional (more complicated) equations for the guiding center approximation (here in the zero-gyroradius limit).  The manifold viewpoint also allows us to infer the presence of conserved quantities for the particles when there is a symmetry (Killing field) tangent to the field sheets.

A practical benefit of the geometrical approach is that the results immediately apply in any curved spacetime.  We are therefore able to move beyond previous work to include the effects of gravity.  We illustrate the physics using the monopole solutions for pulsar and black hole magnetospheres \cite{Michel1973mar,Blandford.Znajek1977jul}.  On each field line, one imposes boundary conditions for the particle flow at a specific place regarded as the ``gap'' that produces the particles.  The particle flow and parallel electric field are then immediately determined everywhere in the magnetosphere.  We can therefore explore the behavior of particles as a function of gap assumptions.

We find a few different qualitative behaviors.  In the limit of gaps that produce a large number of particles (high plasma multiplicity), the parallel electric field vanishes and we recover prior results for particle flow in the force-free limit of ideal magnetohydrodynamics (MHD).  For less efficient gaps there is non-zero (but small) parallel electric field that contributes comparably to the centrifugal ``ExB'' acceleration.  In some finely tuned cases there is significant particle acceleration at the light cylinder, even in this nearly force-free regime.  For some gap parameters, the gravitational attraction overcomes electromagnetic acceleration and particles cannot reach past a given radius; these gaps are too ``weak'' to support the assumed force-free magnetosphere.  In this way one can place restrictions on the allowed gap physics for a given force-free solution, helping identify circumstances in which the force-free approximation will break down globally.

Outside of the MHD limit, our model involves counter-streaming species and hence is susceptible to the two-stream instability.  However, we are expecting a limited presence of these instabilities in realistic magnetospheres. This is because the outflow of the pair plasma from the gap region is moderately hot, and two-stream instabilities are easily suppressed if thermal spreads of the beams exceed their velocities \citep[e.g.,][]{ONeil1968}. In this case, we still expect the bulk velocities of plasma species to be reasonably well described by our two-fluid model.

This paper is organized as follows.  In Sec.~\ref{sec:particle-motion} we review the notion of field sheets and use them to reformulate the zero-gyroradius guiding center approximation of particle motion.  In Sec.~\ref{sec:two-fluid} we study the two-fluid model in the force-free limit, explaining how the global flow can be determined from boundary conditions at gaps.  In Sec.~\ref{sec:magnetospheres} we apply the method to pulsar and black hole magnetospheres with radial field lines.  In Sec.~\ref{sec:consistency} we discuss how the approximation method could be extended to higher orders in perturbation theory.  We conclude in Sec.~\ref{sec:outlook} by mentioning some interesting future directions.  A series of appendices provides supporting material.

Our conventions are as follows.  Latin indices $a,b.c\cdots$ are abstract tensor indices, while Greek indices $\mu,\nu,\cdots$ represent tensor components and run from 0 to 3. Our metric $g_{ab}$ has signature $(-,+,+,+)$.  We use Heaviside-Lorentz units and set $c=G=1$.  Our field strength tensor has $F_{xy}=+B_z$.

\section{Particle  motion on field sheets}\label{sec:particle-motion}

\subsection{Degenerate field and field sheets}
We begin with a quick review of the geometrical approach to degenerate Maxwell fields \cite{Uchida1997aug,Uchida1997oct,Uchida1998jun,Gralla.Jacobson2014dec}, which also establishes our notation.  The field tensor $F_{ab}$ is said to be degenerate and magnetically dominated if
\begin{align}
    \tilde{F}_{ab}F^{ab}=0, \qquad F_{ab} F^{ab} > 0,
\end{align}
where $\tilde{F}_{ab}=\frac{1}{2}\epsilon_{abcd}F^{cd}$ is the Hodge dual of $F_{ab}$.  These conditions are equivalent to $\vec{E}\cdot\vec{B}=0$ and $\vec{B}^2>\vec{E}^2$, respectively.  

Using $B_{0}$ and $E_{0}$ defined as the magnetic and electric fields in the frame where there is no perpendicular electric field in the direction of the magnetic field (Appendix \ref{app: field eigenframe}), the conditions can also be expressed as 
\begin{align}
    E_0 = 0, \qquad B_0 > 0.
\end{align}

Using the property that $F_{ab}$ is a closed two-form (i.e., $dF=0$), it can be shown that the vector space annihilating $F$ is involutive; thus, due to the Frobenius theorem, this vector space is completely integrable, indicating the existence of integrable submanifolds.  For magnetically dominated fields the submanifolds are timelike and may be interpreted as the 2D world sheets swept by the magnetic field lines, or ``field sheets''  \citep{Uchida1997aug,Gralla.Jacobson2014dec}.  We can then say that timelike world lines everywhere tangent to the field sheets represent trajectories of test particles sliding along the magnetic field lines.  That is, if a four-velocity $u^a$ satisfies
\begin{align}\label{stuck}
    F_{ab} u^b = 0,
\end{align}
we say that the particle is \textit{stuck to the field line}.  Conversely, if there exist timelike vectors $u^a$ satisfying \eqref{stuck}, then the field strength is degenerate and magnetically dominated.

Since the field sheets are the submanifolds in the 4D manifold, the spacetime metric $g_{ab}$ induces a metric $h_{ab}$ on each field sheet.  
This ``field sheet metric'' acts on an arbitrary vector as a projection operator onto the field sheet. 
In this context it may be expressed as (Appendix \ref{app: field eigenframe}) 
\begin{align}
    \label{eq: field sheet metric}
    h_{ab}=g_{ab}+\frac{1}{B_{0}^{2}}F_{ac}F\indices{^{c}_{b}}
    =\frac{1}{B_{0}^{2}}\tilde{F}_{ac}\tilde{F}\indices{^{c}_{b}}.
\end{align}
Indeed, when $h_{ab}$ acts on vectors that are tangent to the field sheet, $h_{ab}$ satisfies the usual properties of the metric (e.g. non-degenerate and symmetric). 
For instance, for a vector $v^{a}$ tangent to the field sheet (i.e. $F_{ab}v^{b}=0$), we have 
\begin{align}
    \label{eq:h property1}
    h_{ab}v^{b}=g_{ab}v^{b}=v_{a}.
\end{align}
For each field sheet, there exists a volume form $\epsilon_{ab}$ compatible with the field sheet metric. 
Considering the fact that the field sheet volume form acts as an antisymmetric projection operator that satisfies $\epsilon^{ab}\epsilon_{ab}=-2$, it is given by 
\begin{align}
    \label{eq: field sheet volume form}
    \epsilon_{ab}=\frac{1}{B_{0}}\tilde{F}_{ab}.
\end{align}
This expression together with Eq.~(\ref{eq: field sheet metric}) immediately implies that 
\begin{align}
    \label{eq: h and epsilon}
      h^{ab}={\epsilon}\indices{^a_{c}}{\epsilon}\indices{^{cb}}.
\end{align}
The field sheet metric $h_{ab}$ and the volume form $\epsilon_{ab}$ project vectors onto the field sheet. 
Similarly, it is possible to project vectors onto a local vector space that is perpendicular to the field sheet. 
In general, these vector spaces are not integrable, i.e., they are not the tangent bundle of a submanifold.  
However, one can always consider the local perpendicular plane that is orthogonal to the field sheet. 
Then, the symmetric and antisymmetric projectors $h_{\perp ab}$ and $\epsilon_{\perp ab}$ onto the local perpendicular plane are given by
\begin{align}
    \label{eq: perp plane metric}
    h_{\perp ab}&=g_{ab}-\frac{\tilde{F}_{ac}\tilde{F}\indices{^{c}_{b}}}{B_{0}^{2}}
    =-\frac{1}{B_{0}^{2}}F_{ac}F\indices{^{c}_{b}},
    \\
    \label{eq: perp plane volume form}
    \epsilon_{\perp ab}&=\frac{1}{B_{0}}F_{ab}. 
\end{align}
Similar to Eq.~(\ref{eq: h and epsilon}), one immediately finds
\begin{align}
    \label{eq: hperp and epsilonperp}
    h_{\perp}^{ab}=-{\epsilon_{\perp}}\indices{^a_{c}}{\epsilon_{\perp}}\indices{^{cb}}. 
\end{align}
From Eqs.~(\ref{eq: perp plane metric}) and (\ref{eq: perp plane volume form}), It is clear that for any vector $v^{a}$ tangent to the field sheet, its projection onto the local plane vanishes (i.e. $h_{\perp ab}v^{b}=0=\epsilon_{\perp ab}v^{b}$). 
In other words, from the direct calculations, we can verify that $h_{ab}, \epsilon_{ab}$ and $h_{\perp ab}, \epsilon_{\perp ab}$ are orthogonal:
\begin{align}
    \label{eq: ortho relations}
    0=h_{ac}h_{\perp}^{cb}=
    \epsilon_{ac}h_{\perp}^{cb}=
    \epsilon_{ac}\epsilon_{\perp}^{cb}=
    \epsilon_{ac}\epsilon_{\perp}^{cb}.
\end{align}

On the field sheet, one can define the covariant derivative $D_{a}$ compatible with the field sheet metric $h_{ab}$ by projecting the covariant derivative of a tensor on the field sheet $T\indices{^{b\cdots}_{c\cdots}}$ onto the field sheet: 
\begin{align}
    \label{eq: field sheet derivative}
    D_{a}T\indices{^{b\cdots}_{c\cdots}}
    &=
    h\indices{^{b}_{b'}}\cdots h\indices{^{c'}_{c}}\cdots h\indices{_{a}^{d}}\nabla_{d}T\indices{^{b'\cdots}_{c'\cdots}}.
\end{align}
This definition of $D_{a}$ provides the unique derivative associated with $h_{ab}$ \citep{Wald1984}. 

Additionally, one can solve Eqs.~(\ref{eq: field sheet volume form}) and (\ref{eq: perp plane volume form}) for $F_{ab}$ and $\tilde{F}_{ab}$ as
\begin{align}
    \label{eq: F=B epsilon perp}
    F_{ab}&=B_{0}\epsilon_{\perp ab},
    \\
     \label{eq: F=B epsilon}
    \tilde{F}_{ab}&=B_{0}\epsilon_{ab}.
\end{align}

\subsection{Perturbations}

We will also consider fields that are perturbed away from a degenerate, magnetically dominated field.  Our notation in perturbation theory will be $\bar{A}=A + \varepsilon \delta A$, i.e., a bar indicates an exact quantity, the lack of a decorator indicates a background quantity, a $\delta$ indicates a first-order perturbation, and $\varepsilon$ is the expansion parameter.  For a general Maxwell field $\bar{F}_{ab}$, one can write (Appendix \ref{app: field eigenframe}) 
\begin{align}
\bar{F}\indices{^a_c}\tilde{\bar{F}}^{cb}=-\bar{E}_{0}\bar{B}_{0}g^{ab},
\end{align}
so the perturbation about a degenerate, magnetically dominated background satisfies
\begin{align}
    \label{eq: delta E0 projection}\delta\tilde{F}\indices{^{ac}}F\indices{_{c}^b}+\tilde{F}\indices{^{ac}}\delta F\indices{_{c}^b}=-\delta E_{0}B_{0}g^{ab}.
\end{align}
After projecting on the field sheet (contracting with $h_a{}^d h_b{}^{e}$), this becomes
\begin{align}
    h_{be} \tilde{F}_{dc} \delta F^{cb} & = - \delta E_0 B_0 h_{de} 
\end{align}
After using \eqref{eq: F=B epsilon} on the LHS and \eqref{eq: h and epsilon} on the RHS, we have
\begin{align}
    h_{be} \epsilon_{dc} \delta F^{cb} & = - \delta E_0 \epsilon_d{}^c \epsilon_{c e}.
\end{align}
Denoting the projection (or \textit{pullback}) of a tensor to the field sheet by $|_{\mathcal{S}}$, this equation says 
\begin{align}\label{eq: pullback is delta E}
    \delta F|_{\mathcal{S}} = - \delta E_0 \epsilon.
\end{align}

\subsection{Guiding center approximation on the field sheet}

We now discuss the guiding center approximation \cite{Kruskal1958mar,Northrop1961jul,Vandervoort1960jul} in field sheet language.  The basic physics of this approximation is that in the presence of a uniform magnetic field, the motion of charged particles can be understood as two distinct motions: gyromotion in the direction perpendicular to the magnetic field line and the uniform velocity motion in the parallel direction.  For curved magnetic field lines, one can still separate these motions as long as the gyroradius is much smaller than the curvature radius of the field lines.  When the separation of those scales (gyroradius and curvature radius) is large enough, the motion of a particle can be approximated by the rapid circular motion around a point called the guiding center, which is shown to mostly move along the magnetic field with a small drift motion that goes off of the magnetic field line.

To derive the guiding center approximation relevant to the situation we are interested in, let us first write the equation of motion of a particle in a general curved spacetime in the presence of the EM field $F$.  For a timelike vector $\bar{u}^{a}$ representing the velocity of a test particle with mass $m$ (i.e., $\bar{u}^{a}\bar{u}_{a}=-1$), the equation of motion is given by
\begin{align}
    \label{eq:1 species EoM}
    \varepsilon \bar{u}^{b}\nabla_{b}\bar{u}^{a}=s\bar{F}\indices{^{a}_{c}}\bar{u}^{c},
\end{align}
with  
\begin{align}
    \label{eq:varepsilon=m/e}
    \varepsilon=\frac{m}{e}>0,
\end{align}
where $e>0$ is the elementary charge and $s$ in Eq.~(\ref{eq:1 species EoM}) is the sign of the charge (e.g., $s=-1$ for electron and $s=+1$ for positron). 
As we stated, the fundamental assumption of the guiding center approximation is that the typical curvature radius of the EM field is much larger than the gyroradius of the particle so that the EM field barely changes during one gyration period \citep{Vandervoort1960jul,Northrop1961jul}. 
This idea can be formulated by formally taking $\varepsilon$ to be small, but keeping the velocity of a particle and the EM field to be finite \citep{Kruskal1958mar}. 
One way to understand the justification of this procedure is as follows: the gyroradius is $v_{\perp} m/(eB)$ where $v_{\perp}$ is the perpendicular velocity component of the particles relative to the magnetic field lines of typical field strength $B$. 
Denoting the typical curvature scale of the field as $R$, the assumption of small gyroradius compared to $R$ is given by 
\begin{align}
\frac{v_{\perp} m}{eBR}=\varepsilon \frac{v_{\perp}}{BR}\ll1.
\end{align}
Since this condition is always satisfied by formally taking the value of $\varepsilon$ to be sufficiently small, we can treat $\varepsilon$ as if it were a dimensionless parameter to represent the small radius relative to the typical length scale of the fields. 

Since the order of Eq.~(\ref{eq:1 species EoM}) is reduced by taking the $\varepsilon\to0$ limit, a power series of $\bar{u}^{a}$ in $\varepsilon$ does not capture the entirety of the approximation.
Instead, the WKB type asymptotic expansion was introduced by \cite{Kruskal1958mar} and it was shown to be the correct asymptotic expansion \citep{Berkowitz.Gardner1959}.
It turned out that the power series expansion corresponds to a special case with zero gyromotion. 
In the presence of a strong magnetic field, it is expected that all the particles lose gyromotion through synchrotron radiation \citep{Landau.Lifshitz1975jan,Cai.Gralla.ea2023sepa} within a short period of time. 
For this reason, it is justified to adopt a power series expansion (i.e., zero-gyration limit) if the background field is strong enough. 
As pointed out by many authors \citep{Kruskal1958mar, Vandervoort1960jul, Northrop1961jul}, the electric field parallel to the magnetic field has to be $\mathcal{O}(\varepsilon)$ order for the consistency of the approximation. 
To track this effect, we expand both the field $F$ and $u$ in a power series as
\begin{align}
     \bar{u}^{a}&=u^{a}+\varepsilon \delta u^{a}+\mathcal{O}(\varepsilon^{2}), \label{expandu}
     \\
      \bar{F}_{ab}&=F_{ab}+\varepsilon \delta F_{ab}+\mathcal{O}(\varepsilon^{2}).\label{expandF}
\end{align} 
From the Lorentz force \eqref{eq:1 species EoM} at zero other, we find
\begin{align}
    \label{eq:1 species EoM 0th}
    F_{ab}u^{b}=0.
\end{align}
This implies that the background field $F$ is degenerate and magnetically dominated and that the background particles are stuck to its field lines.  At first order, we find
\begin{align}
    \label{eq:1 species EoM 1st}
    u^{b}\nabla_{b}u^{a}
    &= sF^{ab}\delta u_{b}+s\delta F^{ab}u_{b}.
\end{align}
Notice that the LHS involves only zeroth order terms, while the RHS contains first order terms.  This mixing of orders occurs because the expansion parameter $\varepsilon$ is already present in the equations of motion (\ref{eq:1 species EoM}).  Notice that $s=\pm 1$ appears as the effective charge-to-mass ratio in perturbation theory, since $\varepsilon$ has been scaled out of our definition of perturbations \eqref{expandu} and \eqref{expandF}.

From \eqref{eq: F=B epsilon perp}, we see that $F_{ab}$ is perpendicular to the field sheet.  We may therefore eliminate the first term in the RHS of \eqref{eq:1 species EoM 1st} by projecting onto the field sheet.  Since $u^a$ is already tangent to the field sheet, the projection of second term is just the Lorentz force on the field sheet, and the projection of the LHS is just the field sheet acceleration $u^b D_b u^a$.  It thus follows immediately that \textit{the background particle equation of motion is the Lorentz force law on the field sheet},
\begin{align}
    u^{c}D_{c}u^{a} =  s (\delta F|_{\mathcal{S}})^{ab} u_b.\label{eq: lorentz force on field sheet}
\end{align}
More explicitly, we may use Eq.~\eqref{eq: pullback is delta E} to write
\begin{align}
         u^{c}D_{c}u^{a} & = -s\delta E_{0}\epsilon^{ab}u_{b}.\label{eq:1 species 2d EoM}
\end{align}
This formulates the zero-gyroradius guiding center approximation intrinsically to the field sheet, giving it a very simple appearance.  The traditional (more complicated) form \cite{Vandervoort1960jul,Bacchini.Ripperda.ea2020nov} is obtained by introducing a 3+1 split (choice of frame) and decomposing the velocity in that frame into components parallel and perpendicular to the magnetic field in that frame.  The various terms involving derivatives of the magnetic field arise from the Christoffel symbols associated with the expression of the intrinsic derivative $D_c$ in terms of these external structures.  In the presence of gravity, gravitational force terms also arise from the Christoffel symbols; the intrinsic field sheet metric captures both effects.  While the more complicated formulation is necessary for global numerical simulations involving the guiding center approximation on all field lines at once, the intrinsic formulation has advantages for analytic calculations (as we shall see) and provides a clear geometric interpretation of the motion of particles.

We can go further by using the fact that the field sheet is two-dimensional.  First note that since $u^{a}u_{a}=-1$,  one can antisymmetrize the derivative of $u$ in Eq.~(\ref{eq:1 species 2d EoM}) as
\begin{align}
    \label{eq:1 species 2d EoM 2}
     2 u^{c}D_{[c}u_{a]}&=-s \delta E_{0}\epsilon_{ab}u^{b},
\end{align}
where we use normalized antisymmetrization $A_{[ab]}=(A_{ab}-A_{ba})/2$.  Since every two-form on a two-dimensional manifold is proportional to the volume element, we have $D_{[a}u_{b]}=A \epsilon_{ab}$ for some $A$. Plugging into \eqref{eq:1 species 2d EoM 2} then reveals that $A=s \delta E_0/2$,
\begin{align}\label{eq: d2u is E0}
2 D_{[a}u_{b]} = s \delta E_0 \epsilon_{ab}.
\end{align}
The LHS is just $d_2 u^\flat$, where $d_2$ is the exterior derivative on the field sheet and $u^\flat$ is the one-form $u_a$.  As already noted, the RHS is just $-s$ times the pullback of $\delta F = d \delta A$ to the field sheet~\eqref{eq: pullback is delta E},
\begin{align}\label{duflat}
d_2 u^\flat = -s \delta F|_{\mathcal{S}} = -s (d \delta A)|_{\mathcal{S}},
\end{align}
where in the last step we introduce the perturbation of the vector potential (regarded as a one-form),
\begin{align}
    \bar{A}_{a}=A_{a}+\varepsilon \delta A_{a}+\mathcal{O}(\varepsilon^{2}).
\end{align}
Since the exterior derivative commutes with pullback, we may write \eqref{duflat} as
\begin{align}
    \label{eq: canonical closed}
    d_2\left(u^{\flat} + s\delta A|_{\mathcal{S}}\right)=0,
\end{align}
which says that the canonical momentum is conserved as a one-form on the field sheet.  

Finally, we note that contracting \eqref{eq: d2u is E0} with $\epsilon^{ab}$ gives a formula for $\delta E_0$,
\begin{align}
    \label{eq: 1 species 2d EoM component}
    -s\delta E_{0} = \epsilon^{ab}D_{a}u_{b}= *_2d_2u^\flat,
\end{align}
where $*_2$ is the Hodge dual on the field sheet.

\section{Two-fluid model in the FF limit}\label{sec:two-fluid}
We now consider a cold two-fluid electron-positron plasma. The electron $(-)$ and positron $(+)$ four-velocities are denoted $\bar{u}_{\pm}^{a}$ (satisfying $\bar{u}_{\pm}^{a}\bar{u}_{\pm a}=-1$) and their rest frame number densities are $\bar{n}_{\pm}$.  The electric current is thus given by
\begin{align}
    \label{eq: current=plasma flow}
    \bar{j}^{a}=e\bar{n}_{+}\bar{u}_{+}^{a}-e\bar{n}_{-}\bar{u}_{-}^{a}.
\end{align}
In the two-fluid model, each species obeys the continuity equation and Lorentz force law in the self-consistent electromagnetic field,
    \begin{align}
    \label{eq:EoM}
    \varepsilon \bar{u}_{\pm}^{b}\nabla_{b}\bar{u}_{\pm}^{a}
    &=\pm \bar{F}^{ab}\bar{u}_{\pm b},
    \\
    \label{eq:continuity}
    \nabla_{a}(\bar{n}_{a}\bar{u}_{\pm}^{a})&=0,
    \\
    \label{eq:Maxwell1}
    \nabla_{b}\bar{F}^{ab}&=\bar{j}^{a},
    \\
    \label{eq:Maxwell2}
    \nabla_{a}\bar{\tilde{F}}^{ab}&=0.
\end{align}
where $\varepsilon =m_{e}/e>0$ with $m_{e}$ the mass of the electron.  Eq.~\eqref{eq:EoM} is identical to Eq.~(\ref{eq:varepsilon=m/e}), choosing $s=-1$ for electrons and $s=+1$ for positrons.  

Since our interest is in the regime where the fields are strong and the FFE description is valid, we can assume that $\bar{u}_{\pm}^{a}, \bar{n}_{\pm}, \bar{F}_{ab}$ is expanded as a power series in $\varepsilon$ in a similar manner to the guiding center approximation. 
Namely,
\begin{align}
    \bar{u}_{\pm}^{a}&= u^{a}_{\pm}+\varepsilon \delta u_{\pm}^{a}+\mathcal{O}
    (\varepsilon^{2}),
    \\
    \bar{n}_{\pm}&= n_{\pm}+\varepsilon \delta n_{\pm}+\mathcal{O}(\varepsilon^{2}),
    \\
    \bar{F}_{ab} &= F_{ab}+\varepsilon \delta F_{ab}+\mathcal{O}(\varepsilon^{2}).
\end{align}
To zeroth order in $\varepsilon$, the plasma densities and velocities satisfy 
\begin{align}
     \label{eq:EoM 0th}
    F^{ab}u_{\pm b}&=0,
    \\
    \label{eq:continuity 0th}
    \nabla_{a}(n_{\pm}u_{\pm}^{a})&=0.
\end{align}
That is, each fluid is stuck to field lines and obeys the four-dimensional continuity equation.  The Maxwell equations take the usual form
\begin{align}
    \label{eq:Maxwell1 0th}
    \nabla_{b}F^{ab}&=j^{a}
    \\
     \label{eq:Maxwell2 0th}
    \nabla_{a}\tilde{F}^{ab}&=0
\end{align}
with the zeroth order electric current density provided by the zeroth order plasma flow,
\begin{align}
    \label{eq: current=plasma flow 0th}
    j^{a}&=en_{+}u_{+}^{a}-en_{-}u_{-}^{a}.
\end{align}
It follows from \eqref{eq:EoM 0th} and \eqref{eq: current=plasma flow 0th} that $F_{ab} j^a=0$, i.e., the plasma is force-free.  To emphasize the decoupling of the particle degrees of freedom, we write this as
\begin{align}
    \label{eq:FFE system}
     F_{ab}\nabla_{c}F^{bc}=0.
\end{align}
This force-free description is expected since the small mass limit implies $T_{EM}\gg T_{\mathrm{matter}}$ as long as particles are not ultra-relativistic.  To summarize, the leading order description is force-free electrodynamics with both species stuck to field lines.

Next, let us consider the next order in the perturbation expansion.  The equation of motion \eqref{eq:EoM} becomes
\begin{align}
    \label{eq:EoM 1st}
    u^{b}_{\pm}\nabla_{b}u^{a}_{\pm}
    &= \pm F^{ab}\delta u_{\pm b}\pm\delta F^{ab}u_{\pm b},
\end{align}
which is identical to the expression \eqref{eq:1 species EoM 1st} arising in the guiding center approximation.  In particular, all of the equations \eqref{eq: lorentz force on field sheet}--\eqref{eq: 1 species 2d EoM component} of that approximation apply to each species $u_\pm$, choosing $s=\pm$.  In particular, each species obeys the Lorentz force law on the field sheet, and the canonical momentum of each species is closed as a one-form.  

The electric field may be reconstructed from either species via \eqref{eq: 1 species 2d EoM component},
\begin{align}
    \label{eq: u and E0}
    \delta E_{0} = \mp\epsilon_{ab}D^{a}u^{b}_\pm.
\end{align}
Adding together the $+$ and $-$ equations eliminates the electromagnetic field, giving an equation relating the two species,
\begin{align}
    \label{eq: consistency}
     \epsilon_{ab}D^{a}u_{+}^{b}
     + \epsilon_{ab}D^{a}u_{-}^{b}
     =0
\end{align}
Equivalently, the sum of the fluid four-velocities is closed as a one-form, 
\begin{align}
    \label{eq: consistency d form}
    d_2(u^{\flat}_{+}+u^{\flat}_{-})=0.
\end{align}

Let us now check that there are the correct number of equations to determine the first-order plasma motion ($n_{\pm}$ and $u_{\pm}^{a}$).  Since both species are stuck to field lines \eqref{eq:EoM 0th}, each four-velocity contains only one free component (say, the velocity along the field line), arising from the equations $F_{ab}u_\pm^b=0$ and $u_\pm^a u_{\pm a} = -1$.  The densities $n_\pm$ provide two more quantities, for a total of four to be determined.  An important observation is that the electric current $j^a$ is already known from the force-free background via $j^a=\nabla_b F^{ab}$, and is automatically conserved ($\nabla_a j^a=0$).  Taking the divergence of \eqref{eq: current=plasma flow 0th} shows that $\nabla_{a}(n_{+}u_{+}^{a})=\nabla_{a}(n_{-}u_{-}^{a})$, i.e., the two continuity equations \eqref{eq:continuity 0th} are not independent and count only as a single equation.  We then see that there are four equations: one independent continuity equation, the two components of the requirement \eqref{eq: current=plasma flow 0th} that the particles produce the needed force-free current, and the consistency condition \eqref{eq: consistency} that both species feel the same electric field.  Thus there are the correct number of equations to determine the plasma properties $u_\pm$ and $n_\pm$.

Once the zeroth-order plasma properties are determined, the first-order parallel electric field $\delta E_{0}$ can be calculated from Eq.~\eqref{eq: u and E0}.  It is initially surprising that part of the first-order electromagnetic field can be calculated without solving the first-order field equations.  However, this situation already arises at zeroth order, since all force-free solutions are degenerate ($E_{0}=0$).  Thus we can determine the background parallel electric field (of zero) without explicitly solving the background field equations.  As explored in more detail in Sec. \ref{sec:consistency} below, this kind of staggering is a fundamental feature of the perturbation expansion needed to consider nearly force-free solutions.

So far we have written the equation of motion intrinsically to the field sheet, but we have not done the same for the continuity equation.  This may be done by considering the fact that $u_{\pm}$ is stuck to the field lines:
\begin{align}
     0=\nabla_{a}(n_{\pm}u_{\pm}^{a})&=
     \nabla_{a}\left(h\indices{^{a}_{b}}n_{\pm}u^{b}_{\pm}\right)\notag
     \\
     &=
      \nabla_{a}\left(\epsilon^{ac}\epsilon\indices{_{cb}}n_{\pm}u^{b}_{\pm}\right)\notag
     \\
     &=
    \nabla_{a}\left( \tilde{F}^{ac}\frac{1}{B_{0}}\epsilon\indices{_{cb}}n_{\pm}u^{b}_{\pm}\right)\notag
     \\
     &=
     \tilde{F}^{ac}\nabla_{a}\left(\frac{1}{B_{0}}\epsilon\indices{_{cb}}n_{\pm}u^{b}_{\pm}\right)\notag
     \\
     &=
      B_{0}\epsilon^{ac}D_{a}\left(\frac{1}{B_{0}}\epsilon\indices{_{cb}}n_{\pm}u^{b}_{\pm}\right)\notag
      \\
      &=
      B_{0}D_{a}\left(\frac{1}{B_{0}}\epsilon^{ac}\epsilon\indices{_{cb}}n_{\pm}u^{b}_{\pm}\right)\notag
     \\
     &=
      B_{0}D_{a}\left(\frac{1}{B_{0}}n_{\pm}u^{a}_{\pm}\right), \label{eq: almost 2D continuity}
\end{align}
where we use Eqs.~\eqref{eq: h and epsilon}, \eqref{eq: field sheet volume form} and \eqref{eq:Maxwell2 0th} together with the fact that the species are stuck to the field sheet, $u^a_\pm h_{ab}=u_{\pm b}$.  In fact, this equation holds for any vector stuck to the field sheet: 
\begin{align}\label{eq: field sheet divergence}
\nabla_a v^a = B_0 D_a \left( \frac{v^a}{B_0} \right),
\end{align}
provided only that $F_{ab}v^b=0$ and magnetic flux is conserved, $\nabla_a \tilde{F}^{ab}=0$.  Recalling the formula $\nabla_a V^a =  |g|^{-1/2} \partial_\mu (|g|^{1/2} V^\mu)$ for the divergence of any vector $V$ in a metric with determinant $g$, we see that $B_0$ functions as a kind of volume correction factor, indicating how much transverse area needs to be attached to the field sheet volume element in order to produce the proper spacetime divergence of a vector field tangent to the field sheets.  This geometrical interpretation stems ultimately from the conservation of magnetic flux: to compute a spacetime divergence from intrinsic field sheet quantities, one needs to know how the field sheets themselves diverge in spacetime, which is encoded in the magnetic field strength by conservation of magnetic flux.

Returning to Eq.~\eqref{eq: almost 2D continuity}, we see that the four-dimensional continuity equation can be written in two-dimensional form as
\begin{align}
    \label{eq: 2D continuity}
    D_{a}J_\pm ^a =0, \qquad J_\pm^a \equiv \frac{n_{\pm}u_{\pm}^{a}}{B_{0}}.
\end{align}
That is, the conserved current on the field sheet is the particle number current per unit magnetic field strength.  

We may form a useful mental picture by regarding each magnetic field line as carrying a unit $\Phi_0$ of magnetic flux.  Geometrically, each field becomes a variable-thickness tube that widens/narrows where the magnetic field decreases/increases, keeping the flux equal to some given $\Phi_0$.  At any given point along the tube, the cross-sectional area is $B_0/\Phi_0$ by definition, so $n_\pm \Phi_0 / B_0$ is the linear density of particles in the flux tube.  The constant factor $\Phi_0$ is just an arbitrary choice of units (size of flux tube) expressing the fact that $\lambda_\pm=n_\pm/B_0$ is proportional to the linear density of particles.  We therefore regard $J_\pm^a=\lambda_\pm u_\pm^a$ as the conserved particle current of each species on each 1+1 dimensional field sheet, keeping in mind that the sheets are infinitesimally thickened according to the rule of constant magnetic flux.

\subsection{Symmetry and conserved quantities}

The geometrical perspective allows us to efficiently infer the existence of conserved quantities when there is sufficient symmetry.  In particular, suppose that the system possesses a symmetry tangent to the field sheets.  That is, suppose there is a vector field $\chi$ such that
\begin{align}\label{eq: symmetry}
    F_{ab} \chi^a=0, \quad \mathcal{L}_\chi h = \mathcal{L}_\chi u_\pm = \mathcal{L}_\chi n_\pm = 0,
\end{align}
where $\mathcal{L}$ is the Lie derivative.  Notice that we do not require $\chi$ to be a symmetry of the spacetime; it need only be a symmetry of each field sheet and the fields defined on it.  We will make use of Cartan's formula for the Lie derivative of a form,
\begin{align}\label{eq: cartan}
    \mathcal{L}_{v}\omega =v\cdot d \omega + d(v\cdot\omega)
\end{align}
where the notation $v \cdot \omega$ indicates contraction of a vector $v$ with the first index of a form $\omega$.  

Dotting \eqref{eq: consistency d form} with $\chi$ and using Cartan's formula \eqref{eq: cartan} on the field sheet manifold together with the vanishing of the Lie derivative \eqref{eq: symmetry} shows immediately that
\begin{align}
    d_{2} \left(u_{+}^{b}\chi_{b}+u_{-}^{b}\chi_{b}\right)=0.
\end{align}
The quantity in parentheses is thus constant everywhere on the field sheet manifold.  We will use the symbol $-\Gamma$ for this conserved quantity,
\begin{align}
    \label{eq: cnsrv1}
   \Gamma = - u_{+}^{b}\chi_{b}-u_{-}^{b}\chi_{b}.
\end{align}

For the rotating magnetospheres we primarily study, the Killing field is $\chi = \partial_t + \Omega \partial_\phi$, where $\Omega$ is a constant on each field sheet (but can differ from sheet to sheet).  In this case we have $-u^a_\pm \chi_a=E_\pm-\Omega L_\pm$, where $E_\pm$ and $L_\pm$ are the lab-frame energy and angular momentum per unit rest mass of a fictitious particle following the relevant flow.  Thus $\Gamma$ is some kind of ``total'' energy minus angular momentum.  We emphasize that $\Gamma$ is not only constant along the flow but in fact constant everywhere on the field sheet.\footnote{The constancy on the flow can be understood in terms of four-dimensional conserved quantities: Each species has a separate conserved quantity $p^a_{\pm} \chi_a$ as long as the canonical momentum $p_\pm^a=m u^a\pm A^a$ is constructed from a vector potential sharing the symmetry, and $\Gamma$ is the sum of these conserved quantities.  However, the existence of a vector potential sharing the symmetry is not required for $\Gamma$ to be conserved.}

We can also find conserved quantities associated with the conservation of particle current (continuity equation).  
In fact, any conserved current $J^a$ on a two-dimensional manifold gives rise to a conserved quantity $\epsilon_{ab}J^a \chi^b$ when $\chi$ is a symmetry ($\mathcal{L}_\chi J = \mathcal{L}_\chi h = 0)$.  To see this, we consider Cartan's formula acting on $J\cdot \epsilon$,
\begin{align}
    \mathcal{L}_\chi (J \cdot \epsilon) = \chi \cdot d_2(J \cdot \epsilon) + d_2(\chi\cdot J \cdot \epsilon).
\end{align}
The LHS vanishes by the assumption of a symmetry, while the first term on the RHS vanishes by the assumption of a conserved current---the differential forms statement of conservation is $d(J\cdot\epsilon)=0$, equivalent to $D_a J^a=0$. Thus we find that $\chi\cdot J \cdot \epsilon$ is constant on each field sheet.  For the currents $J^{a}=n_{\pm}u^{a}_{\pm}/B_{0}$  of Eq.~\eqref{eq: 2D continuity}, we find conserved quantities
\begin{align}
     \label{eq: cnsrv2}
    C_{\pm} = \epsilon_{ab} J_{\pm}^a \chi^b = \frac{n_{\pm}}{B_{0}}\epsilon_{ab}u_{\pm}^{a}\chi^{b}.
\end{align}
These quantities have clear interpretations for a globally stationary configuration ($\chi=\partial_t$).  Since $J_\pm$ may be viewed as the 1+1 dimensional current density in an infinitesimal flux tube (see discussion below \eqref{eq: 2D continuity}), $C_\pm$ is just the total particle current through the flux tube.  That is, the tube acts like a wire carrying a steady current $C_{\pm}$.  

Alternatively, we may interpret $C_\pm$ in terms of three-vector quantities, as follows.  The magnetic field is expressed as $B^a=\tilde{F}^{ab} \chi_b=B_0 \epsilon^{ab} \chi_b$, or equivalently or $\vec{B}=B_0 \hat{b}$ where $\hat{b}$ is the spatial part of $\epsilon^{ab} \chi_b$.  Every two-vector on the field sheet similarly promotes to a three-vector in the direction $\hat{b}$, where the component is determined by contraction with $\epsilon^{ab} \chi_b$.  In particular, the particle current density is expressed as $\vec{j}_\pm = (n_\pm u^a_\pm) (\epsilon^{ab} \chi_b) \hat{b}$. Thus the conserved quantity expresses the particle three-current density (spatial part of $n_\pm u^a_\pm$) in units of magnetic field,
\begin{align}\label{eq: jB}
    \vec{j}_\pm = C_\pm \vec{B}.
\end{align}
Eq.~\eqref{eq: jB} holds only when $\chi=\partial_t$, but for any timelike $\chi^a$ similar arguments establish that $\vec{j}_\pm = (C_\pm/z) \vec{B}$, where $z$ is the ``redshift factor'' $z=\sqrt{-\chi^a\chi_a}$ and spatial vector components refer to the local frame associated with an observer $U^a=\chi^a/z$ co-moving with the symmetry.

Each of $C_{+}$ and $C_{-}$ is constant on the field sheet, but these quantities are not independent.  Rather, they are related by the requirement \eqref{eq: current=plasma flow 0th} that the flow of charged particles produces the force-free current $j^a$, which is predetermined by solving the force-free equations.  The electric current $j^a$ itself is conserved and tangent to the field sheet, so by \eqref{eq: field sheet divergence} we have a two-dimensional conserved current $D_a(j^a/B_0)=0$, implying a conserved quantity
\begin{align}
    \label{eq: define Cj}
    C&=\frac{\epsilon_{ab}j^{a}\chi^{b}}{B_{0}},
\end{align}
which by \eqref{eq: current=plasma flow 0th} is just the difference 
\begin{align}
    \label{eq: Cj/e=C+-C-}
    C=e(C_{+}-C_{-}).
\end{align}
This quantity has the analogous interpretations already discussed for $C_\pm$: for globally stationary configurations $\chi=\partial_t$ it is the electric current through a flux tube and also the proportionality $\vec{j}=C \vec{B}$ between the electric current and magnetic field.  In this context the conserved quantity is well-known, and it has also been used for rotating force-free magnetospheres (see Appendix~A of \cite{Gralla.Lupsasca.ea2017dec} for details).

It will be convenient to consider the ratio of $C_{\pm}$ to $C$, so we define 
\begin{align}
    \label{eq: define mu}
    \mu_{\pm}=-e\frac{C_{\pm}}{C}=-\frac{en_{\pm}\epsilon_{ab}u_{\pm}^{a}\chi^{b}}{\epsilon_{ab}j^{a}\chi^{b}}.
\end{align}
Since $C_{\pm}$ and $C$ satisfy Eq.~(\ref{eq: Cj/e=C+-C-}), $\mu_{+}$ and $\mu_{-}$ obey the constraint
\begin{align}
    \label{eq: mu--mu+=1}
    \mu_{-}-\mu_{+}=1.
\end{align}
Eq.~\eqref{eq: define mu} can be rearranged as equation for $n_\pm$,
\begin{align}\label{eq: calculate n}
    n_{\pm}=-\mu_{\pm}\frac{\epsilon_{ab}j^{a}\chi^{b}}{e \epsilon_{ab}u^{a}\chi^{b}}.
\end{align}
Finally, by dotting $\chi^a$ into $j^a$ in Eq.~\eqref{eq: current=plasma flow 0th} and using \eqref{eq: calculate n} to eliminate $n_{\pm}$, we and find an equation for $u_{\pm}^{a}$,
\begin{align}
    \label{eq: current req}
    -\frac{j^{a}\chi_{a}}{\epsilon_{ab}j^{a}\chi_{a}}
    &=\mu_{+}\frac{u_{+}^{a}\chi_{a}}{\epsilon_{ab}u_{+}^{a}\chi_{a}}
    -
    \mu_{-}\frac{u_{-}^{a}\chi_{a}}{\epsilon_{ab}u_{-}^{a}\chi_{a}}.
\end{align}

\subsection{Equations for flow}

In the previous section we showed that the presence of a field sheet symmetry $\chi$ entails the existence of two independent conserved quantities, which may be taken to be the triple $(\Gamma,\mu_+,\mu_-)$ obeying the constraint $ \mu_--\mu_+ =1$.  These quantities can determine the flow ($n_\pm$ and $u^a_\pm$) as follows.  Since each flow velocity $u^a_\pm$ has only one independent component (say, the velocity on the field sheet---ultimately this derives from the constraints $u_\pm^a u_{\pm a}=-1$ and $F_{ab}u_\pm^b=0$), the two equations Eqs.~\eqref{eq: cnsrv1} and \eqref{eq: current req} determine both $u_a^+$ and $u^a-$.  These quantities then determine $n_\pm$ via Eq.~\eqref{eq: calculate n}.

To perform this procedure in practice, it is helpful to explicitly impose the constraints $u_\pm^a u_{\pm a}=-1$ and $F_{ab}u_\pm^b$ so that $u^a_+$ and $u^a_-$ are expressed in terms of some scalar functions $\beta_+$ and $\beta_-$.  In the spirit of staying geometrical as long as possible, we will do so by considering a null basis $\{\ell^a,n^a\}$ of the field sheet.  These null directions can also be defined as the eigenvectors of the field strength (its ``principal null directions''---Appendix \ref{app: field eigenframe}), where the eigenvalues vanish for this degenerate field.  They are also eigenvectors of the dual field strength with eigenvalues $\mp B_0$ (\ref{eq:Ftilde PND =B0 PND}), and we use the condition $\tilde{F}^a{}_b \ell^b=-B_0 \ell^a$ to fix them up to normalization.  We further fix the normalization by imposing 
\begin{align}
    \label{eq: PND normalization}
    \ell^{a}n_{a}=-\frac{1}{2},
\end{align}
which means that the four-velocities are given by
\begin{align}
    \label{eq: define beta}
u_{\pm}^{a}=\beta_{\pm}\ell^{a}+\frac{1}{\beta_{\pm}}n^{a},
\end{align}
where $\beta_\pm$ are scalar functions,
\begin{align}
    \beta_\pm =-2 u^a_\pm n_a.
\end{align}

 There remains the gauge freedom to rescale $\ell \to \psi \ell$, $n \to \psi^{-1} n$, and $\beta_\pm \to \psi \beta_\pm$ for any scalar $\psi$.  When the system has a symmetry expressed as a Killing vector field $\chi$, we will fix this freedom by demanding
\begin{align}
    \label{eq: PND gauge fixing}
    \chi_{a}\ell^{a}=-\frac{1}{2}.
\end{align}
Noting $\epsilon_{ab} = 2(\ell_a n_b - \ell_b n_a$) to be consistent with Eq.~(\ref{eq:epsilon lpm=mp lpm}), this condition entails
\begin{align}\label{eq: epsilon gauge fixing}
    \epsilon_{ab} \chi^a \ell^b = \frac{1}{2}.
\end{align}
Another useful relation with this normalization choice is
\begin{align}
    \label{eq: chi^2=2chiell}
    \chi^{a}\chi_{a}&=2\frac{(\ell^{a}\chi_{a})(n^{a}\chi_{a})}{\ell^{a}n_{a}}
    =2n^{a}\chi_{a}.
\end{align}

When decomposing $u_{\pm}$ as a linear combination of $\ell$ and $n$, Eqs.~(\ref{eq: cnsrv1}) and (\ref{eq: current req}) become
\begin{align}
    \label{algebraic eq1}
    \Gamma&=\frac{1}{2}\left(\beta_{+}+\beta_{-}\right)
    -\frac{1}{2}\left(\frac{1}{\beta_{+}}+\frac{1}{\beta_{-}}\right)\chi^2,
    \\
    \label{algebraic eq2}
     -\frac{j^{a}\chi_{a}}{\epsilon_{ab}j^{a}\chi^{b}}
     = & \mu_+\frac{\beta_{+}^{2}-\chi^2}{\beta_{+}^{2}+\chi^2}
    -
    \mu_{-}\frac{\beta_{-}^{2}-\chi^2}{\beta_{-}^{2}+\chi^2}. 
\end{align}
where $\chi^2=\chi^a \chi_a$ was introduced using \eqref{eq: chi^2=2chiell}.  The second equation can be solved for $\beta_+^2$ in terms of $\beta_-^2$, after which the first equation becomes an eigth order polynomial in $\beta_-$.  Thus in general there can be 8 solutions for $\beta_+$ and $\beta_-$.  Physical solutions must be real-valued and satisfy the conditions that (1) $u_{+}^{a}$ and $u_{-}^{a}$ are both future-directed, and (2) the rest frame densities $n_{+}$ and $n_{-}$ are both positive.  

In the definite examples we study in Sec.~\ref{sec:magnetospheres}, we find that there is at most one physical solution for any given choice of conserved quantities.  Once $\beta_+$ and $\beta_-$ are determined, the rest frame densities $n_{\pm}$ can be readily obtained from \eqref{eq: calculate n} using \eqref{eq: define beta},
\begin{align}
    \label{eq: rest density}
    n_{\pm}
      = &-\mu_{\pm}\frac{\epsilon_{ab}j^{a}\chi^{b}}{e \left(
    \beta_{\pm}\ell^a\chi_{a}-\frac{1}{\beta_{\pm}}n^{a}\chi_{a}\right)}.
\end{align}

The parallel electric field can similarly be determined from \eqref{eq: 1 species 2d EoM component} using \eqref{eq: define beta}.  However, a more convenient formula follows from expressing \eqref{eq: 1 species 2d EoM component} as $d_2 u^\flat_\pm=\pm \epsilon \delta E_0$ (equivalent to \eqref{eq: d2u is E0} with $s=\pm$).  Since $\chi^a$ is a symmetry, it follows that $-d_2(\chi \cdot u^\flat_\pm)=\pm \chi \cdot \epsilon \delta E_0$.  Then dotting again with $\ell^a$ and using   $\epsilon_{ab}\chi^a \ell^b=1/2$ \eqref{eq: epsilon gauge fixing} gives the formula
\begin{align}
    \label{eq: parallel E 1st}
    \delta E_{0} & = \mp 2\ell^{a} D_{a}(u_{\pm}^{b}\chi_{b}) \\
    & = \mp 2\ell^{a} D_{a}\left(
    \beta_{\pm}\ell^b\chi_{b}-\frac{1}{\beta_{\pm}}n^{b}\chi_{b}\right).
\end{align}

\subsection{Physical flow condition}\label{sec:turning-point}

For each choice of conserved quantities $\Gamma,\mu_\pm$ at each spacetime point, the properties of Eqs.~(\ref{algebraic eq1}) and (\ref{algebraic eq2}) determine whether there is a physical flow solution.  However, since $\Gamma$ and $\mu_{\pm}$ need to be constant on the field sheet, the existence of a physical flow solution at one point in spacetime does not guarantee the existence of a physical flow solution everywhere in the magnetosphere: there could be ``forbidden regions'' for this choice of conserved quantities.  If we were discussing a single particle, the boundary of the physical region would be a turning point where the velocity momentarily vanishes.  For our fluids obeying the continuity equation, the velocity will again vanish at the boundary of the physical region (otherwise the fluid would continue on), but the density becomes infinite to conserve particle current.  These are not true turning points but rather \textit{singularities} indicating the breakdown of the two-fluid model as a plasma realization for the force-free solution.  The breakdown may be understood physically by reasoning that individual particles ``want'' to turn around at the turning point, but actual turning would entail a flow with two velocities at one point in spacetime, in violation of the fluid assumption.  In light of this implied turning, we will continue to refer to the boundaries of physical regions as turning points.  We will also see that a turning point for one species implies the same turning point for the other species.

Let us now make these statements precise.  From \eqref{eq: cnsrv1}, the positron rest-frame density may be expressed as
\begin{align}
    n_{+}=\frac{C_{+}}{\epsilon_{ab}u^{a}_{+}\chi^{b}}.
\end{align}
Since $C_+$ is constant, the density blows up where $\epsilon_{ab}u^{a}_{+}\chi^{b}=0$, i.e., where $u_+^a$ and $\chi^a$ become collinear.  This can only happen when $\chi^a$ is timelike.  In this case $\chi^a$ provides a time orientation for the portion of the field sheet manifold where it is timelike, and the orthogonal vector $\epsilon_{ab}\chi^b$ provides a space orientation.  We can thus interpret the vanishing of $\epsilon_{ab}u^{a}_{+}\chi^{b}$ as a moment of zero spatial velocity, where a single particle would change direction.  This provides the ``turning point'' interpretation of the singularity in $n_+$, where the flow velocity becomes collinear with the Killing field.  

Analogous statements hold for electrons: they have turning points of diverging $n_-$ where $u^a_-$ and $\chi^a$ are co-linear.  However, the turning points of both species must occur at the same physical points in spacetime so that the force-free current $j^a=e( n_{+}u^a_{+}- n_{-}u^a_{-})$ \eqref{eq: current=plasma flow 0th} is finite.  That is, if there is a turning point for one species, there will be a turning point for both. At such a point, both four-velocities are parallel to the Killing field,

\begin{align}
    u_{+}^{a}=u_{-}^{a}=\frac{\chi^{a}}{\sqrt{-\chi^{b}\chi_{b}}}.
\end{align} 
Plugging these expressions for $u_{+}$ and $u_{-}$ in Eq.~(\ref{eq: cnsrv1}), we find that the condition 
\begin{align}
    \label{eq: turning point condition}
    \left(\frac{\Gamma}{2}\right)^{2}+\chi^{a}\chi_{a}=0
\end{align}
is satisfied at the turning point.  Note that when the value of $\Gamma/2$ is exactly equal to the global minimum of $\chi^{a}\chi_{a}$, there is only one point that satisfies Eq.~(\ref{eq: turning point condition}). 
Traditionally, the 2-dimensional surface on which $-\chi^{a}\chi_{a}$ takes the global minimum is usually referred to as a separation surface or the stagnation surface \citep{Takahashi.Nitta.ea1990nov,Hirotani:2006gu,Hirotani.Pu2016feb}.

Eq.~\eqref{eq: turning point condition} is satisfied at the boundaries of physical flow regions.  This implies that the quantity on the LHS has a definite sign in any physical region.  We have shown in four separate cases that the sign is positive: (1) at large $r$ for a rotating magnetosphere in an asymptotically flat spacetime (since $\chi=\partial_t + \partial_\phi$ so that $\chi^a \chi_a\to +\infty$ in the asymptotic region); (2) near any turning point (by solving Eqs.~(\ref{algebraic eq1}) and (\ref{algebraic eq2}) perturbatively); (3) in the high-multiplicity (MHD) limit (see Eq.~\eqref{eq:FFMHD beta0} below); (4) in every numerical example we study.  We therefore strongly suspect---but have not  proven---that at each spacetime point and for each choice of conserved quantities, a necessary and sufficient condition for the existence of a physical flow solution is
\begin{align}
    \label{eq: global flow condition}
    \left(\frac{\Gamma}{2}\right)^{2}+\chi^{a}\chi_{a}\geq 0.
\end{align}

\subsection{Force-free ideal MHD limit}\label{sec:FFMHD limit}
The small mass expansion of the two-fluid model describes the behavior of two species of plasma in the force-free background. 
When the system has an abundance of electrons and positrons, it is expected that their velocity and density become almost identical. 
This can be understood as even a very small difference in the velocity and density between electrons and positrons contributing to the finite strength of the electric current $j^{a}$ due to the availability of many electrons and positrons. 
Since this limit corresponds to the ideal FF limit of the cold MHD system, we call this the force-free MHD (FFMHD) limit.

In the presence of field sheet symmetry, the FFMHD limit can be obtained by taking either one of $\mu_{\pm}$ to be large, i.e., by expanding in  $1/\mu_{\pm}$.  Since these quantities obey $-\mu_++\mu_-=1$ \eqref{eq: mu--mu+=1}, taking one to be large implies that the other is large as well, with $\mu_+ \sim \mu_-$  in the limit. 
Thus increasing $\mu_{\pm}$ leads to a progressively more balanced flow of positrons and electrons (nearly equal amount of electrons and positrons flowing), as desired.  

We will also keep the first correction in $1/\mu_{\pm}$ since it provides useful analytical expressions for checking numerical results.  By solving Eqs.~(\ref{algebraic eq1}) and (\ref{algebraic eq2}) up to first order in $1/\mu_{-}$ in our normalization with \eqref{eq: chi^2=2chiell}, we find
\begin{align}
  \beta_{+}&=\beta_{0}+\frac{1}{\mu_{-}}\beta_{+}^{(1)}+\mathcal{O}\left(\frac{1}{\mu_{-}^{2}}\right),
    \\
    \beta_{-}&=\beta_{0}+\frac{1}{\mu_{-}}\beta_{-}^{(1)}+\mathcal{O}\left(\frac{1}{\mu_{-}^{2}}\right),
\end{align}
where 
\begin{align}
    \label{eq:FFMHD beta0}
    \beta_{0}&=
    \frac{\Gamma}{2}\pm\sqrt{\left(\frac{\Gamma}{2}\right)^{2}+\chi^2},
    \\
    \label{eq:FFMHD beta1}
    \beta_{+}^{(1)}=-\beta_{-}^{(1)}
    &=
    \frac{\left(\chi^2+\beta_{0}^2\right)
   \left((A-1) \chi^2+(A+1) \beta_{0}^2\right)}{8 \chi^2 \beta_{0}},
\end{align}
with 
\begin{align}
    A=-\frac{j^{a}\chi_{a}}{\epsilon_{ab}j^{a}\chi^{b}}.
\end{align}
In this limit, one can verify that 
\begin{align}
    \chi_{a}u_{0}^{a}=-\frac{\Gamma}{2}
\end{align} 
with $u_{0}$ being the leading bulk velocity,
\begin{align}
    u^a_{0}=\beta_{0}\ell^a+\frac{1}{\beta_{0}}n^a.
\end{align} 
The conservation of $\chi_{a}u^{a}_{0}$ in the FFMHD limit is discussed in \citep{Lovelace.Mehanian.ea1986sep,Contopoulos1995jun,Contopoulos.Kazanas.ea1999jan,Petrova2015jan,Petrova2017may} in the 3+1 formulation in flat spacetime. 
In particular, Ref.~\cite{Contopoulos.Kazanas.ea1999jan} used this conserved quantity to estimate the energy of the particle wind in the dipole FF solution. 
The same observation was made by \cite{Gralla.Jacobson2014dec} in the context of the curved spacetime, assuming that the plasma follows geodesics on the field sheet.  
Our formulation clarifies that the conservation of $\chi_{a}u_{0}^{a}$ is justified only in the FFMHD limit where there is an abundance of electrons and positrons.  

Notice that the solution \eqref{eq:FFMHD beta0} for $\beta_0$ exists if and only if the condition \eqref{eq: global flow condition} is satisfied, as remarked above that equation.

Since we derive FFMHD as the large $\mu_{\pm}$ limit of the FF two-fluid model, it is possible to find an expression for the first-order parallel electric field $\delta E_{0}$. 
Using Eq.~(\ref{eq: parallel E 1st}), we find
\begin{widetext}
\begin{align}
     \label{eq: FFMHD E0}
      \delta E_{0}& =
    \frac{1}{\mu_{-}}\left[
    \frac{-\chi^2 \left(\beta_{0} ^2+\chi^2\right)^3 A'+(\chi^{2})' A \left(\beta_{0} ^2+\chi^2\right)
   \left(\beta_{0} ^4+\chi^4-4 \beta_{0} ^2 \chi^2\right)-(\chi^{2})' \left(\chi^2-\beta_{0} ^2\right)^3}{16 \beta_{0}
   ^3 \chi^4}\right]
   +
   \mathcal{O}\left(\frac{1}{\mu_-^2}\right),
\end{align}
\end{widetext}
where the prime is understood as the derivative  $-2\ell^a D_{a}$. 
To derive this expression, we first take the derivative of Eqs.~(\ref{algebraic eq1}) and (\ref{algebraic eq2}) along the field sheet ($D_{a}$ derivative). 
This operation leads to two linear simultaneous equations for $D_{a}\beta_{+}$ and $D_{a}\beta_{-}$. 
By computing Eq.~(\ref{eq: parallel E 1st}) and replacing $D_{a}\beta_{\pm}$ with the solutions to these equations, one can obtain the expression for $\delta E_{0}$. 

Notice from Eq.~(\ref{eq: FFMHD E0}) that $\delta E_{0}$ scales as $1/\mu_{-}$.  This corresponds to the vanishing of the parallel electric field in the ideal MHD limit of $\mu_{-}\to\infty$.

\section{Plasma flow in FF magnetospheres}\label{sec:magnetospheres}

In the previous section we gave a general treatment of the force-free limit of two-fluid theory.  In particular, we showed that a field sheet symmetry gives rise to two conserved quantities on each field sheet, and that the values of these conserved quantities fully determine the plasma flow and parallel electric field.   In a physical model, the values of the conserved quantities should be determined by boundary conditions at edges of the force-free regions.  

In pulsar and black hole magnetospheres, the relevant edges are gap regions that produce the pairs filling the magnetosphere.  The gaps inject particles into the magnetosphere, whose properties (velocity and density) determine the conserved quantities in the bulk force-free magnetosphere.  In principle, one should use gap microphysics to determine a prescription for the electron and positron injection velocity and density as functions of the compact object properties and background force-free solution.  In this paper we will instead be agnostic to the gap microphysics and simply explore the various plasma behaviors that are possible, varying the injection densities and velocities.  In order to keep the treatment analytic, we consider pulsar and black hole magnetospheres with radial poloidal field lines (monopole magnetospheres).

\subsection{Pulsar magnetosphere}\label{sec:pulsar magnetosphere}

We will model the pulsar as a spherically symmetric, conducting star.  The metric outside of the star is the Schwarzschild metric,
\begin{align}
    ds^{2}=-\alpha^{2} dt^{2}+\frac{1}{\alpha^{2}}dr^{2}+r^{2}d\theta^{2}+r^{2}\sin^{2}{\theta}d\phi^{2}
\end{align}
where 
\begin{align}
    \alpha=\sqrt{1-\frac{2M}{r}}.
\end{align}
The force-free solution with radial poloidal field lines is \cite{Michel1973mar,Blandford.Znajek1977jul,Lyutikov:2011tq,Gralla.Jacobson2014dec}
\begin{align}\label{Fmichel}
    F=q \sin{\theta}d\theta\wedge \left(d\phi- \Omega dt + \Omega \frac{dr}{\alpha^{2}}\right),
\end{align}
where $q$ and $\Omega$ are constants.\footnote{Eq.~\eqref{Fmichel} remains a solution when the angular velocity $\Omega$ depends on $\theta$, and there are simple generalizations with a time-dependent angular velocity vector and/or a moving star \cite{Brennan.Gralla.ea2013sep,Gralla.Jacobson2014dec,Brennan:2013ppa}.  The analysis of this paper could be straightforwardly repeated for those solutions.}  An observer rotating with angular velocity $\Omega$ sees no electric field, i.e., $F_{ab} U^b=0$ for $U \propto \partial_t + \Omega \partial_\phi$. Thus the solution satisfies the conducting boundary condition for a rotating star (of any radius).  It is therefore regarded as the magnetosphere of a rotating, conducting star that has been magnetized with radial field lines.

The constant $q$ encodes the magnetic charge of the solution.  In a physical magnetosphere model, the solution would be ``split'' by adjoining an overall factor of $\textrm{sign}(\cos \theta)$, making it reflection-symmetric about the equator and removing the magnetic charge.  (This also introduces an equatorial current sheet.)  Rather than writing out these factors, we will simply work in the northern hemisphere, $0<\theta<\pi/2$.  In this context it is more physical to regard $q=B_{*}R_{*}^{2}$ as a characteristic magnetic flux for a star of radius $R_*$ with surface field strength $B_*$.  We take $B_*>0$ (hence $q>0$) for definiteness.

By writing the solution \eqref{Fmichel} as $F=-q d(\cos \theta)\wedge d(\phi-\Omega (t-r_*)$, where $r_*(r)$ is the ``tortoise coordinate'', we see that the field sheets correspond to the surfaces where $\cos \theta$ and $\phi-\Omega(t-r_*)$ are constants.  Different values of $\phi-\Omega(t-r_*)$ just correspond to changing $\phi$ (rotating about the symmetry axis of the solution), so only the value of $\cos \theta$ has physical significance.  We can think of the field sheets as labeled by their polar angle $\theta$.  We will speak of each value of $\theta$ as a single field sheet, or equivalently a (radial) poloidal field line.

The solution \eqref{Fmichel} is stationary and axisymmetric, with Killing vectors $\partial_{t}$ and $\partial_{\varphi}$.  As already remarked, the linear combination 
\begin{align}
    \chi=\partial_{t}+\Omega \partial_{\phi}
\end{align}
is tangent to the field sheets ($F_{ab} \chi^b=0)$, so this is the field sheet Killing vector.

From \eqref{Fmichel}, the four-current is found to be 
\begin{align}
    \label{eq: mono j}
    j^{a}=\nabla_{b}F^{ab}=-\frac{2q\Omega \cos{\theta}}{r^{2}}\left[\frac{1}{\alpha^{2}}(\partial_{t})^{a}+(\partial_{r})^{a}\right],
\end{align}
which is a null vector (i.e., $j^{a}j_{a}=0$). Then, we find that 
\begin{align}
    \label{eq: mono jchi}
    j^{a}\chi_{a}=\frac{2q\Omega \cos{\theta}}{r^{2}}=\epsilon_{ab}j^{a}\chi^{b}.
\end{align} 
Note that the equality $j^a \chi_a=\epsilon_{ab}j^a \chi^b$ is special to the case under consideration.  Choosing the outgoing null direction for $\ell$, the principal null directions in our normalization (\ref{eq: PND normalization}) and (\ref{eq: PND gauge fixing}) are
\begin{align}
    \label{eq:mono l+}
    \ell &=\frac{1}{2\alpha^{2}}\partial_{t}+\frac{1}{2}\partial_{r},
    \\
    \label{eq:mono l-}
    n&=
     \frac{1+\frac{1}{\alpha^{2}}\Omega^{2}r^{2}\sin^{2}{\theta}}{2}\partial_{t}
     + \frac{-\alpha^{2}+\Omega^{2}r^{2}\sin^{2}{\theta}}{2}\partial_{r}
     +\Omega \partial_{\phi}.
\end{align}
Notice that we have
\begin{align}
    \label{eq: Michel mono chi^2}
    \chi^{a}\chi_{a}=2 n^{a}\chi_{a}=-\alpha^{2}+\Omega^{2}r^{2}\sin^{2}{\theta}.
\end{align}
Since each field line runs over all values of $r$, the existence of global solutions requires Eq.~(\ref{eq: global flow condition}) to be satisfied everywhere.  This in turn implies the bound
\begin{align}
    \label{eq:schd mono glabal condition}
    \Gamma&\geq2\sqrt{ 1-3\left(M\Omega\sin{\theta}\right)^{\frac{2}{3}}},
\end{align}
taking into account that $\Gamma$ \eqref{eq: cnsrv1} must be positive since $\chi^a$ is timelike in portions of the magnetosphere.

For each choice of the conserved quantities $\Gamma$ and $\mu_\pm$ on each field sheet, we may solve Eqs.~(\ref{algebraic eq1}) and (\ref{algebraic eq2}) with 
Eqs.~(\ref{eq: mono jchi}) and (\ref{eq: Michel mono chi^2}) to find the plasma velocity and density distributions everywhere on the field sheet.  Since $\mu_\pm$ are not independent (satisfying $\mu_- - \mu_+=1$), there are two conserved quantities to choose per field sheet.  Since the physics of gaps at the stellar surface ultimately determines the values of these conserved quantities, we will parameterize the possible choices by two physical quantities expressed in the frame co-rotating with the stellar surface.  Any two quantities would suffice; we choose the positron Lorentz factor and multiplicity, denoted $\gamma_{+}^*$ and  $\lambda_{+}^*$, respectively.  In terms of the co-rotating frame four-velocity
\begin{align}\label{Ua}
    U^{a}=\frac{\chi^{a}}{\sqrt{-\chi^{b}\chi_{b}}},
\end{align}
the positron Lorentz factor and multiplicity are given by 
\begin{align}
    \label{eq:boundary gammap}
    \gamma_{+}^*&=-u^{a}_{+}U_{a}|_{R_{*}} \\
        \label{eq:boundary lambda}
    \lambda_{+}^*&=-\frac{\left.n_{+}u_{+}^{a}U_{a}\right|_{R_{*}}}{n^*_{\rm GJ}} = \frac{\gamma^*_+ n_+|_{R_*}}{n^*_{\rm GJ}},
\end{align}
where $n^*_{\rm GJ}$ (the co-rotating Goldreich-Julian density) is defined as
\begin{align}
    n^*_{\rm GJ}=\left|\frac{j^{a}U_{a}}{e}\right|_{R_*} = \frac{2 \frac{\Omega B_*}{e} \cos \theta}{ \sqrt{1 - \frac{2M}{R_*} - \Omega^2 R_*^2 \sin^2 \theta}}.
\end{align}
In second equality we use \eqref{eq: mono jchi} together with $q=B_* R_*^2$.  The absolute value signs disappear in light of our restriction to the northern hemisphere (equivalently consideration of the split monopole) and choice of positive magnetic field $B_*>0$.  With these choices the charge density at the surface is negative $(j^a U_a > 0)$, and the positron multiplicity ranges between $0$ (no positrons) to $\infty$ (many positrons).  The Goldreich-Julian number density is the minimum particle density needed to realize the force-free solution, i.e., the electron density if there are no positrons.  However, the no-positron limit is not physically achievable for this null-current solution, since the electrons would have to move at light speed.  Nevertheless, physical quantities that are finite in the limit $\lambda_+ \to 0$ can be used as approximations for the situation with very few positrons and very large electron Lorentz factor.

\begin{figure*}[ht]
\centering
  \begin{minipage}[b]{0.45\linewidth}
  \centering
    \includegraphics[keepaspectratio, scale=0.55]{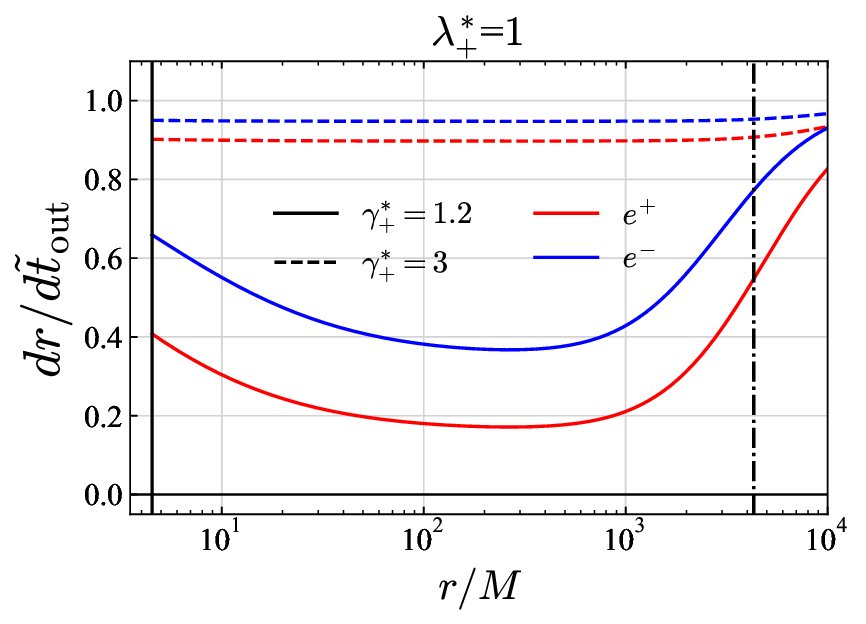}
    \subcaption{}
    \label{fig: Michel r velocity e p (a)}
  \end{minipage}
  \hfill
  \begin{minipage}[b]{0.45\linewidth}
  \centering
    \includegraphics[keepaspectratio, scale=0.55]{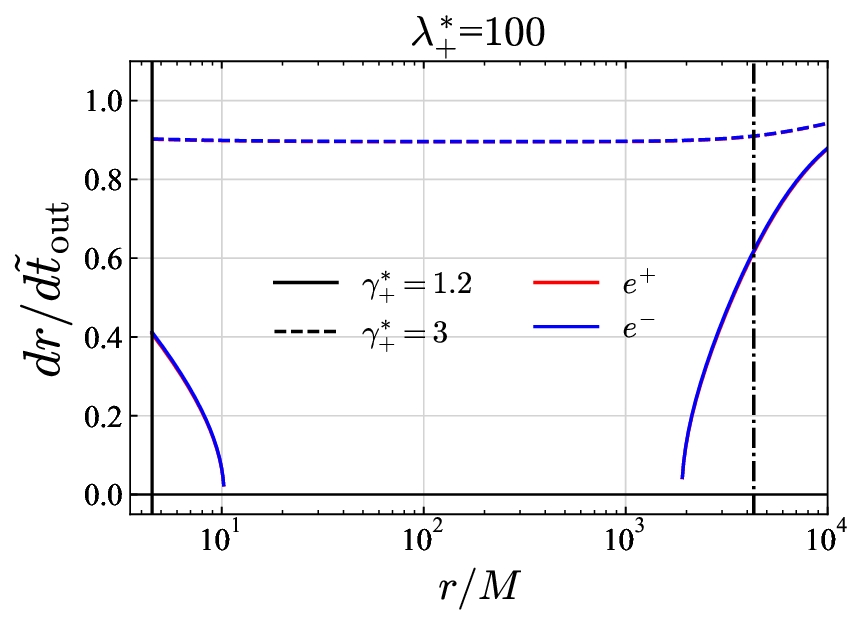}
    \subcaption{}
    \label{fig: Michel r velocity e p (b)}
  \end{minipage}
  \caption{\justifying{The radial velocity $dr/d\tilde{t}_{\rm out}$ of electrons (blue) and positrons (red) as a function of positron Lorentz factor $\gamma^*_+$ and multiplicity $\lambda^*_+$ at the surface of the star. The time coordinate $\tilde{t}_{\rm out}$ is the outgoing Eddington-Finkelstein time defined by Eq.~(\ref{eq: outgoing EF time}). In these plots, the speed of light is given by $dr/d\tilde{t}_{\rm out}=1$. In the low-multiplicity case (left panel) the two species have different velocities, while in the high-multiplicity case (right panel) they have nearly identical velocities (MHD-like behavior).  For the solid-line trajectories on the right, the particles have insufficient energy to overcome the star's gravitational pull.  These parameter choices do not allow the plasma to fill the magnetosphere and therefore cannot realize the force-free solution.   The star parameters are $M=1.5 M_{\odot}$, $R_{*}=10\text{km}$, and $P=0.1$s, and we consider the field line $\theta=\pi/6$.  The light cylinder is shown as a vertical dashed line.}}
  \label{fig: Michel r velocity e p}
\end{figure*}

A choice of $\gamma^*_+$ and $\lambda^*_+$ at a given value of $\theta$ specifies how the gap loads particles onto the corresponding field sheet. For each choice, the field sheet conserved quantities $\Gamma$ and $\mu_{+}$ are computed by eliminating $n_{+}$ from Eq.~(\ref{eq:boundary lambda}) using Eq.~(\ref{eq: rest density}) and then simultaneously solving Eqs. \eqref{eq: mu--mu+=1}, (\ref{algebraic eq2}), (\ref{eq:boundary gammap}) and (\ref{eq:boundary lambda}) for $\mu_{+}, \beta_{\pm}|_{R_{*}}$. 
Given the value of $\beta_{\pm}|_{R_{*}}$, $\Gamma$ is then determined by Eq.~(\ref{algebraic eq1}). 
In general, there are two sets of ($\Gamma,\mu_{+}$) that have the same $\gamma_{+}^*$ and $\lambda_{+}^*$. 
 These correspond to ingoing and outgoing plasma flow; we choose the outgoing branch.

Before presenting results, let us establish some physical intuition resulting from the interplay of particle conservation and the requirement that the force-free current is realized.  We can think of the given force-free charge density $\rho$ and field-aligned current density $J$ as fixed targets that must be met by the electron and positron fluids via the equations $\rho=e(\rho_+-\rho_-)$ and $J=e(\rho_+ v_ +-\rho_+ v_{-}$).  (Here we deal with lab frame quantities and discuss only the field-aligned component of the current.)  We focus on a small region of space where the target current is roughly constant.  If one of the fluids accelerates significantly over this region, then the other must ``respond'' so as to keep the charge-current fixed.  Over the small region, the conservation of particle number is just the constancy of $\rho_{\pm} v_{\pm}$, so the densities of the species scale as $\rho_\pm \sim c_{\pm}/v_\pm$ for some constant $c_{\pm}$.  In order to keep $\rho=e(\rho_+-\rho_-)$ roughly constant, we conclude that
\begin{align}\label{rough-scaling}
\frac{c_+}{v_+} - \frac{c_-}{v_-} \sim \rm{const}.
\end{align}
In particular, acceleration of one species must be accompanied by acceleration of the other species.  Changes in density are similarly compensated via the scaling $\rho_\pm \sim c_{\pm}/v_\pm$.

We therefore conclude that the two species tend to ``pull each other along''. This intuition is markedly different from electric-field induced acceleration, which would accelerate the species in opposite directions.  In the nearly force-free regime we consider, the electric-field acceleration contributes comparably to other forces, such that the overall effect follows the intuition that the species will roughly move in tandem.  The general behavior of particles in a rotating magnetosphere can be understood as a combination of three effects: outward centrifugal acceleration from being stuck to rotating field lines, inward gravitational attraction from the mass of the star, and the tendency for the two species to move together.

We will now show some examples.  Since the plasma is confined to field lines, the velocity along the field lines is the important quantity to understand.  However, there is some arbitrariness in choosing what quantity to consider.  The coordinate radial velocity $dr/dt$ is conventional in flat spacetime, but can be misleading near the star because position-dependent time dilation effects can mask the ``true'' acceleration effects.  In the limit of a compact star (black hole limit), these effects become dominant, such that all causal trajectories near the surface have $dr/dt$ nearly equal to zero.

To avoid this difficulty, we define a new time coordinate $d\tilde{t}_{\rm out}$ with the property that outgoing null rays have unit coordinate velocity, $dr/d\tilde{t}_{\rm out}=1$.  The needed definition is
\begin{align}
    \label{eq: outgoing EF time}
    \tilde{t}_{\rm out}=t-2M\log{\left|\frac{r-2M}{M}\right|},
\end{align}
which is also the outgoing Kerr-Schild coordinate in the non-spinning case $a=0$ (Appendix \ref{app: Kerr-Schild coords}).\footnote{Note that this time coordinate was first considered by Eddington \cite{Eddington:1924pmh} and Finkelstein \cite{Finkelstein1958may}, although it is the null version that today bears their name.}  The radial velocity $d r/d\tilde{t}_{\rm out}$ expresses outgoing particle speed in a range of $0$ to $1$, as in the familiar flat spacetime radial velocity.  The relationship to the four-velocity components is given in Eq.~\eqref{drdtKS} below.

Fig.~\ref{fig: Michel r velocity e p} shows the radial velocity $dr/d\tilde{t}_{\rm out}$ of electrons and positrons with varying choices of density and multiplicity at the star. We take the pulsar parameters to be $M=1.5M_{\odot}$, $R_{*}=10\text{km}$, $P=2\pi/\Omega=0.1 \text{s}$ where $M_{\odot}$ is the mass of the sun and consider the field sheet $\theta=\pi/6$.  For modest multiplicity (Fig.~\ref{fig: Michel r velocity e p (a)}), the electrons and positrons have distinct velocities, but their velocity curves have a similar shape, in line with the intuition that they move in tandem.  When the multiplicity is large (Fig.~\ref{fig: Michel r velocity e p (b)}), the species move on almost exactly the same trajectory, a reflection of the FFMHD limit discussed in Sec.~\ref{sec:FFMHD limit}.  In this plot we also show an example where the particles would fail to escape the gravitational pull of the star and cannot actually realize the force-free solution (see Sec.~\ref{sec:turning-point} for further discussion).

\begin{figure}[t]
\centering
    \includegraphics[keepaspectratio, scale=0.5]{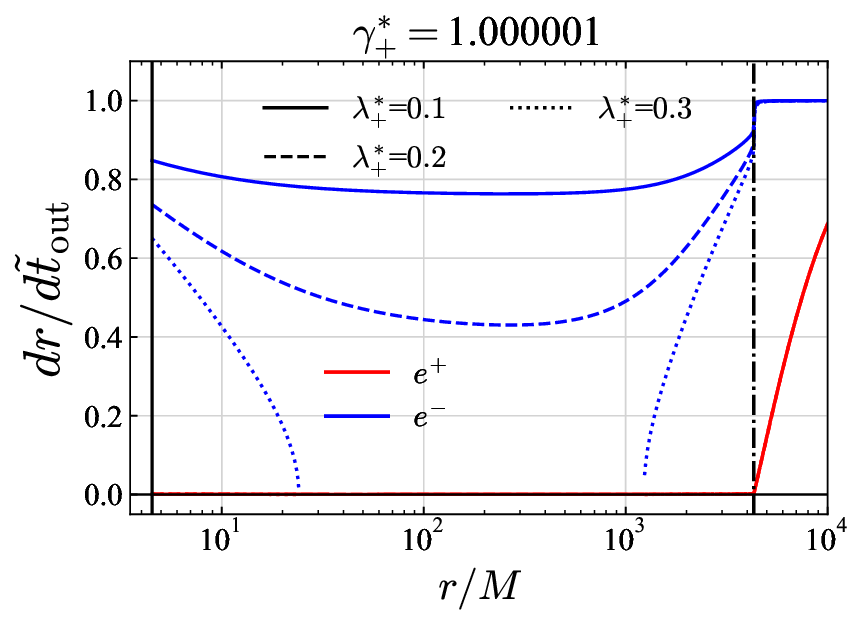}
  \caption{
  \justifying Light cylinder acceleration in a finely tuned region of parameter space.  Centrifugal acceleration of the low-velocity positrons causes extreme acceleration of the modest-velocity to electrons (see text for details).  The setup and parameters are the same as Fig.~\ref{fig: Michel r velocity e p} except where indicated. 
  \label{fig: Michel r velocity e p SA}}
\end{figure}

Fig.~\ref{fig: Michel r velocity e p SA} illustrates some more exotic behavior that occurs for low multiplicity ($\lambda_+^* \lesssim 1$) and extremely low initial radial velocity ($\gamma_{+}^*\sim 1$).  The positrons remain at low radial velocity until the light cylinder, outside of which they must accelerate to remain stuck to the field lines.  The electrons must respond with their own acceleration to maintain the approximate constancy of $c_+/v_+ + c_-/v_-$ [Eq.~ \eqref{rough-scaling}].  Since $v_+$ is small, a small change in $v_+$ entails a large change in $c_+/v_+$, which (assuming $c_+ \sim c_-$) must be compensated by a large change in $1/v_-$.  But since $v_-$ is not small, this large change in $1/v_-$ requires a large change in $v_-$ itself, visible as a nearly vertical segment of the blue curve in Fig.~\ref{fig: Michel r velocity e p SA}.  This extreme light cylinder acceleration was studied previously in Ref.~\cite{Scharlemann1974oct} in flat spacetime; here we include gravity, which prevents the particles from reaching the light cylinder when the initial velocity is too low (dotted line in the figure).

\begin{figure*}[ht]
\centering
   \begin{minipage}[b]{0.45\linewidth}
    \includegraphics[keepaspectratio, scale=0.5]{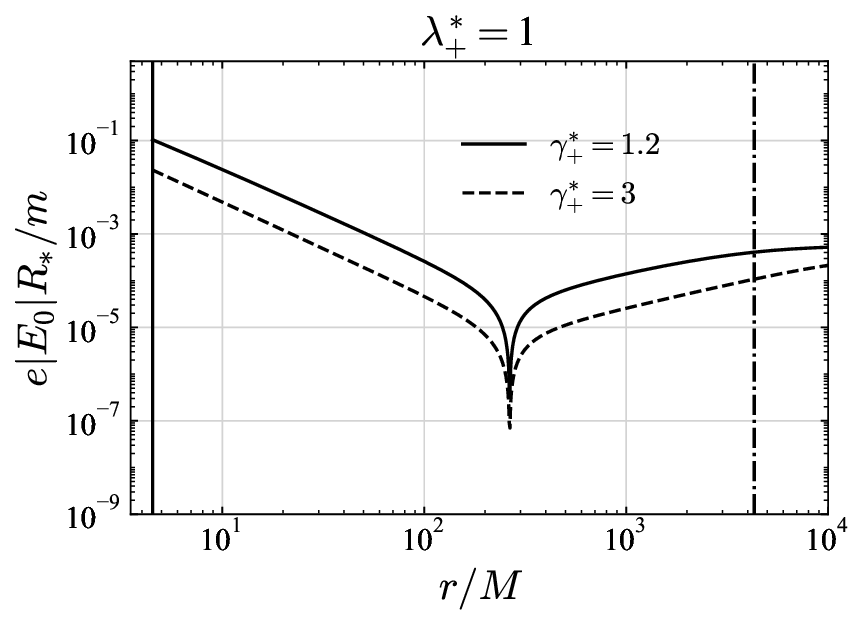}
  \end{minipage}
  \hfill
   \begin{minipage}[b]{0.45\linewidth}
    \includegraphics[keepaspectratio, scale=0.5]{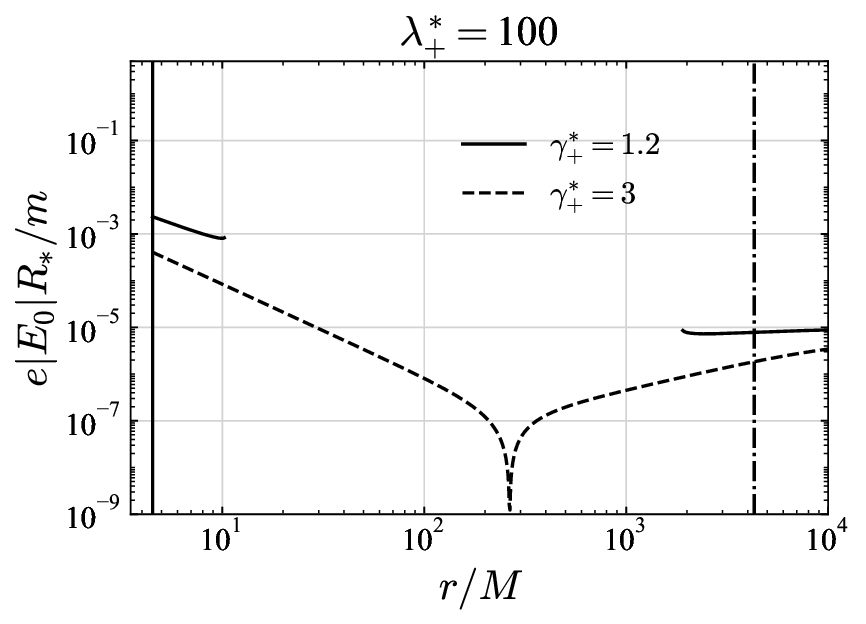}
    \end{minipage}
  \caption{
 \justifying{The parallel electric field of the flow in Fig.~\ref{fig: Michel r velocity e p}. The sign of $\delta E_{0}$ is positive at the star and becomes negative before reaching the LC. 
 The strength of the electric field is larger for the flow that has a larger separation in the velocity of electrons and positrons. 
 Notice that even at the turning points, the electric field does not diverge. }
  \label{fig: Michel E0}
  }
\end{figure*}

\begin{figure}[ht]
\centering
    \includegraphics[keepaspectratio, scale=0.5]{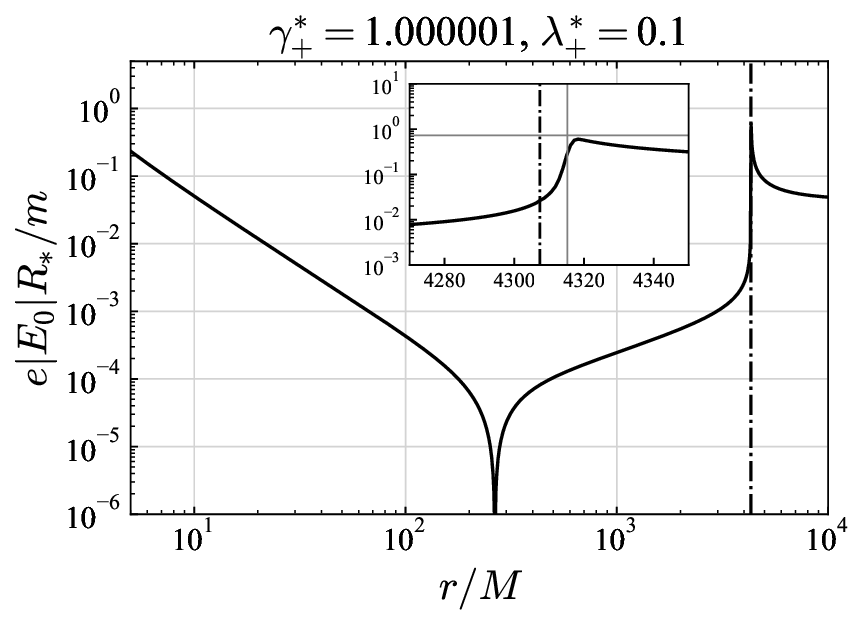}
  \caption{
 \justifying{The parallel electric field for the flow of Fig.~\ref{fig: Michel r velocity e p SA}, where features light cylinder acceleration.  
 Our analytic estimates (\ref{eq:location SA}) and (\ref{eq: strongest E at LC}) for location and value (respectively) of the strongest electric field are shown as gray lines in the inset.}}
  \label{fig: Michel E0 SA}
\end{figure}

Figs.~\ref{fig: Michel E0} and \ref{fig: Michel E0 SA} show the parallel electric field \eqref{eq: parallel E 1st} for Figs.~\ref{fig: Michel r velocity e p} and \ref{fig: Michel r velocity e p SA}, respectively.  Since the parallel electric field is the only force capable of accelerating the species in opposite directions, its size tracks the disparity in acceleration between the species.  The largest disparity (and hence largest electric field) occurs at light cylinder acceleration.  As shown in the inset and derived analytically in appendix~\ref{app: strong E at LC}, the electric field is strongest just outside the light cylinder.

Large electric fields could invalidate our solution in two ways.  First, if $e |E_0|M \gtrsim m$, then particles could gain energy comparable to their rest mass over a gravitational radius, and one would worry about triggering pair production (though realistically this would occur only for $e |E_0|M \gg m$).  Second, if $|E_0| \gtrsim B_0$, then our perturbation theory breaks down and the nearly force-free two-fluid solution cannot be trusted.  The parameters chosen for the figures are far from these regimes.   Indeed, even the fine-tuned light cylinder acceleration involves a modest electric field of only $e |E_0|M / m \approx .1$, which corresponds to $|E_0|/B_0 \approx .1 m/(e M B_0) \approx 10^{-16}$ for typical pulsar parameters.  It is therefore possible in principle for light cylinder acceleration to occur in nature.

\subsection{Black hole magnetosphere}

We now turn to black hole magnetospheres with radial field lines, using the approximate analytic solution found by Blandford and Znajek \cite{Blandford.Znajek1977jul}. Though the magnetic field configuration is almost identical to the case of a rotating star, there are some important differences.  First, the four-current is spacelike as opposed to null.  The black hole solution therefore requires two species of particle to realize it physically, even in the ultrarelativistic limit.

Another key difference is that instead of a single light cylinder we now have two light surfaces, the inner light surface (ILS) and outer light surface (OLS) \cite{Komissarov2004may,Gralla.Jacobson2014dec}. The OLS is conceptually identical to the usual light cylinder in the pulsar magnetosphere: outside of the OLS, the field lines rotate too quickly for a co-rotating observer to be timelike, and particles must move outward.  The ILS is an inverted version: inside of the ILS the field lines rotate too \textit{slowly} for a co-rotating observer to be timelike (they aren't keeping up with the dragging of inertial frames), and particles must move \textit{inward}.  This means that the source of plasma (gap region or regions) must lie between the two light cylinders, and there will be both inward and outward flows.

The BZ solution is an approximate solution valid for a small spin.  The mathematics of the small spin expansion has subtleties on account of the fact that, as the spin approaches zero, the OLS approaches infinity, while the ILS approaches the event horizon.  The naive expansion where the spin approaches zero at fixed coordinate position thus only covers a region between the light surfaces, far from both.  The original BZ work was able to calculate the leading energy extraction rate using a matching procedure without explicit consideration of additional expansions, but this method breaks down at higher orders \citep{Tanabe:2008wm}.  A complete treatment requires additional limits to resolving the OLS and ILS \citep{Armas.Cai.ea2020apr}, although only the OLS is necessary to compute the monopole energy extraction rate \citep{Camilloni.Dias.ea2022jul}.

Since the light surfaces play a key role in particle motion, a complete treatment resolving all possible phenomena would likely require explicit consideration of all three limits.  For example, with sufficient fine tuning of gap physics, we expect that the ILS should be able to accelerate particles inwards towards the black hole, just as the OLS can accelerate them outwards to infinity (as we found in the prior section).  Studying this phenomenon would require a limit that distinguishes the ILS from the event horizon, i.e., the ``near'' limit of \cite{Armas.Cai.ea2020apr}.  However, for simplicity we will consider only generic boundary conditions for which ILS-specific behavior should not arise, an expectation that is verified \textit{ex-post facto} by confirming that perturbative quantities remain small.  In effect, we work perturbatively far from the ILS, but extend quantities through it by continuity.  By contrast, we keep terms that distinguish the OLS from infinity.

We begin with the BZ solution in the form given in \cite{Gralla.Jacobson2014dec},
\begin{align}\label{FmichelBZ}
    F=q \sin{\theta}d\theta\wedge \left(d\phi- \frac{a}{8 M^2} \left( dt - \frac{dr }{\alpha^2} \right) \right).
\end{align}
This field strength satisfies the force-free equations in the Kerr metric perturbatively assuming $a \ll M$.  It is identical to Eq.~(\ref{Fmichel}) with the identification
\begin{align}
    \Omega=\frac{a}{8M^{2}},
\end{align}
i.e., the BZ solution is just a Michel monopole in the Kerr background that rotates with angular velocity $a/8M^{2}$ (which is half the horizon angular velocity).  The field sheet Killing vector field is simply
\begin{align}
    \label{eq:BZ chi}
    \chi=\partial_{t}+\frac{a}{8M^{2}}\partial_{\phi}.
\end{align}

The light surfaces are determined by solving $\chi^2=0$.  Since the field strength \eqref{FmichelBZ} is linear in the spin, it is tempting to also consider $\chi^2$ to this order.  However, one finds $\chi^2 =\alpha^2 + \mathcal{O}(a^2)$ and there are no solutions for the light surfaces. This is the already-mentioned failure of the naive limit to resolve the light surfaces. To understand the physics better, we may expand $\chi^2$ to second order using the exact Kerr metric (\ref{eq: Kerr metric}):
\begin{align}
    \label{eq:BZ chi^2 2nd order}
    \chi^{2}
    &=-\alpha^{2}+\left(\frac{a}{8M^{2}}\right)^{2}r^{2}\sin^{2}{\theta}\notag
    \\
    &
    -\left(
    \frac{2M\cos{\theta}}{r^{3}}+\frac{\sin^{2}{\theta}}{2M r}
    \right)a^{2}+\mathcal{O}(a^{3}).
\end{align}
If we assume that $r \gtrsim M$ then the third term is negligible on account of $a \ll M$, but the second term cannot be dropped because $r$ may be arbitrarily large.  Solving the equation then gives the perturbative location of the OLS.  Similarly, for $r \lesssim M$ the second term can be dropped (but not the third), and solving the equation gives the perturbative location of the ILS.  The results are:
\begin{align}
    \label{eq: BZ OLS}
    r_{\rm{OLS}}&=\frac{8M^{2}}{a\sin{\theta}}\\
        \label{eq: BZ ILS}
    r_{\rm{ILS}}-2M&=\left(\frac{1}{2}-\frac{\sin^{2}{\theta}}{8}\right)\frac{a^{2}}{M}.
\end{align}
As $a\to0$, the OLS approaches infinity, while the ILS coincides with the horizon $r=2M$.

As described above, we will work perturbatively away from the ILS, ignoring any special (finely tuned) phenomena associated with that surface.  We thus make the assumption $r \gtrsim M$ and drop the third term in \eqref{eq:BZ chi^2 2nd order}.  This makes the expression for $\chi^2$ identical to the Schwarzschild monopole case (\ref{eq: Michel mono chi^2}),
\begin{align}
    \label{eq:BZ chi^2}
    \chi^{2}=-\alpha^{2}+\left(\frac{a}{8M^{2}}\right)^{2}r^{2}\sin^{2}{\theta}.
\end{align}
By similar reasoning, when computing the principal null directions $\ell$ and $n$ for the BZ solution, we arrive at the same expressions (\ref{eq:mono l+}) and (\ref{eq:mono l-}).

That is, the BZ expressions for $\chi^2$, $\ell$, and $n$ are given by equations \eqref{eq:mono l+}, \eqref{eq:mono l-}, and \eqref{eq: Michel mono chi^2}, respectively, using $\Omega=a/(8M^2)$.  
The current density, however, differs from that of the Schwarzschild monopole. 
Using 
\begin{align}
    j^{\mu}=\nabla_{\nu}F^{\mu\nu}=\frac{1}{\sqrt{-g}}\partial_{\nu}\left(\sqrt{-g}F^{\mu\nu}\right),
\end{align}
and working in the Kerr background to first order in $a$, we find
\begin{align}
    \label{eq:BZ current1}
    j&=-\frac{2q\cos{\theta}}{r^{2}}\left(\frac{a}{8M^{2}}\right)\left[\left(1-\frac{16M^{3}}{r^{3}}\right)\frac{1}{\alpha^{2}}\partial_{t}+\partial_{r}\right]
    \\
    \label{eq:BZ current2}
    &=-\frac{2q\cos{\theta}}{r^{2}}\left(\frac{a}{8M^{2}}\right)\left[
    2\left(1-\frac{8M^{3}}{r^{3}}\right)\ell-\frac{16M^{3}}{r^{3}\alpha^{3}}n
    \right].
\end{align}
Note that the first line and the second line are equal only to first order in $a$. 
Contrary to the Schwarzschild monopole, the current is no longer null. In fact, it is everywhere spacelike $j^{2}>0$, which can be easily confirmed from Eq.~(\ref{eq:BZ current2}).
To first order, the only difference between the Schwarzschild monopole and the BZ monopole is the current, whose difference can be traced back to the frame-dragging effect due to the Kerr background. Using \eqref{eq:BZ current1} and \eqref{eq:BZ chi}, we find
\begin{align}
    \label{eq: BZmono jchi}
    j^{a}\chi_{a}&=\frac{2q \cos{\theta}}{r^{2}}\left(\frac{a}{8M^{2}}\right)\left(1-\frac{16M^{3}}{r^{3}}\right),
    \\
    \label{eq: BZmono epsilonjchi}
    \epsilon_{ab}j^{a}\chi^{b}&=\frac{2q \cos{\theta}}{r^{2}}\left(\frac{a}{8M^{2}}\right).
\end{align} 
With Eqs.~(\ref{eq:BZ chi^2}), (\ref{eq: BZmono jchi}) and (\ref{eq: BZmono epsilonjchi}), we can solve Eqs.~(\ref{algebraic eq1}) and (\ref{algebraic eq2}) to determine plasma flow from a choice of conserved quantities.

\begin{figure*}[ht]
\centering
  \begin{minipage}[b]{0.45\linewidth}
    \includegraphics[keepaspectratio, scale=0.5]{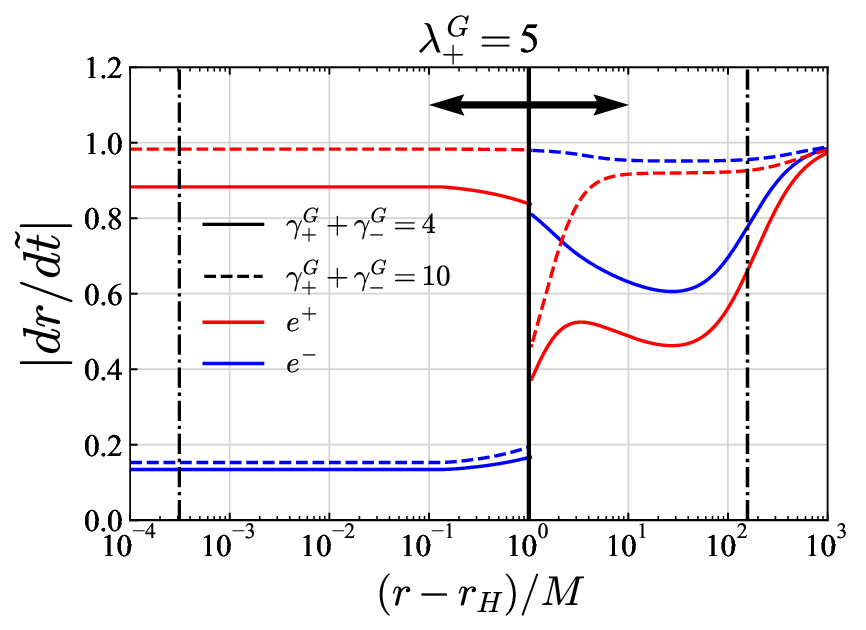}
  \end{minipage}
  \begin{minipage}[b]{0.45\linewidth}
    \includegraphics[keepaspectratio, scale=0.5]{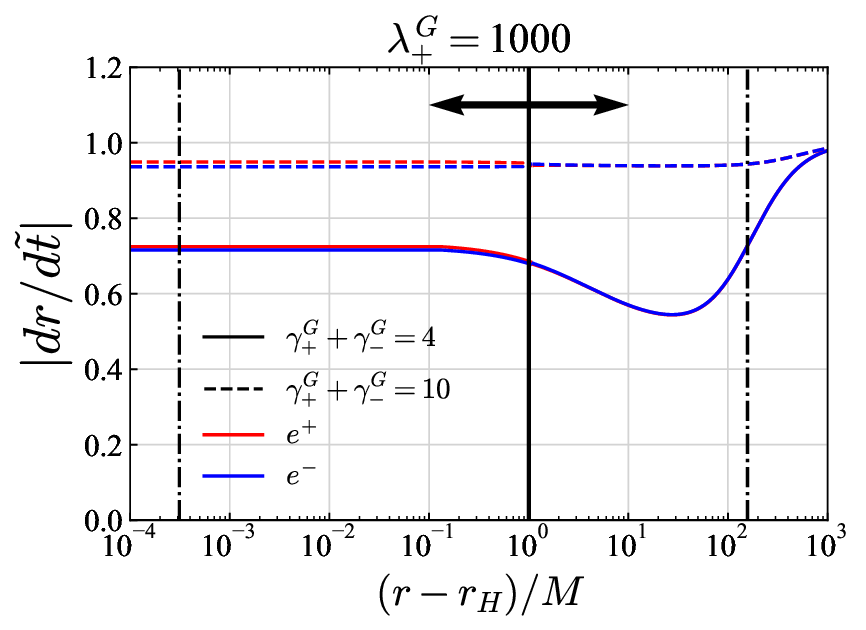}
    \end{minipage}
  \caption{
  \justifying{ Plasma radial velocities for the BZ monopole with $a/M=0.1$ and $\theta=\pi/6$. The gap location is $r_{G}=3M$ and is shown as a solid line.  The inner and outer light surfaces \eqref{eq: BZ ILS} and \eqref{eq: BZ OLS} are shown as dash-dotted lines.  The time coordinate $\tilde{t}$ represents $\tilde{t}_{\rm out}$/$\tilde{t}_{\rm in}$ above/below the gap. (These choices are made so that the radial velocity of outgoing light rays is $dr/d\tilde{t}_{\rm out}=1$ in the outgoing region, while the radial velocity of ingoing light rays is $dr/d\tilde{t}_{\rm in}=-1$ in the ingoing region.)  
  The qualitative behavior is similar to the pulsar magnetosphere: when the multiplicity is large at the gap, the flow reaches the FFMHD limit. }
  \label{fig: BZ velocity}
  }
\end{figure*}

\begin{figure}[ht]
\centering
    \includegraphics[keepaspectratio, scale=0.5]{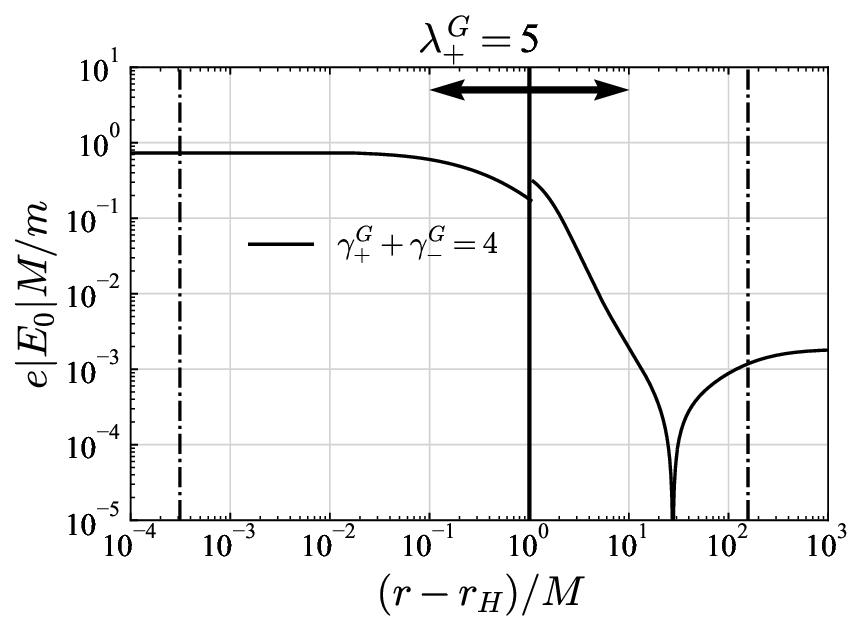}
  \caption{
  \justifying{The parallel electric field associated with the flow of Fig.~\ref{fig: BZ velocity} left. The electric field is discontinuous at the gap region because the gap is two-sided.}
  \label{fig: BZ E0}
  }
\end{figure}

In the pulsar problem we took the gap location to be the stellar surface.  In the black hole problem, there is no natural gap location and simulations suggest different locations \citep{Levinson.Cerutti2018aug,Chen.Yuan2020jun,Crinquand.Cerutti.ea2020apr,Yuan:2025war}.  We assume that the gap is co-rotating with the plasma (with four-velocity $U^a$ \eqref{Ua}) and let its radius $r_G$ be a free parameter.  The gap radius must lie strictly between the light cylinders but is otherwise a free parameter.  To parameterize the particle output we choose the sum of the gamma factors $ \gamma_{+}^{G}+\gamma_{-}^{G}$ of positrons and electrons, and the positron multiplicity $\lambda_{+}|^{G}$.  These are given by
\begin{align}
    \label{eq:boundary gammap BH}
    \gamma_{+}^{G}+\gamma_{-}^{G}&=-(u^{a}_{+}+u^{a}_{-})U_{a}|_{r_{G}}
    \\
    \label{eq:boundary lambda BH}
    \lambda_{+}^{G}&=-\left.\frac{n_{+}u_{+}^{a}U_{a}}{n_{\rm{GJ}}}\right|_{r_{G}},
\end{align}
with $n_{\rm{GJ}}$ is given by 
\begin{align}
    n_{\rm{GJ}}=\left|\frac{j^{a}U_{a}}{e}\right|_{r_G} = \frac{\frac{2q \cos{\theta}}{er_{G}^{2}}\left(\frac{a}{8M^{2}}\right)\left(1-\frac{16M^{3}}{r_{G}^{3}}\right)}{\sqrt{1-\frac{2M}{r_{G}}-\left(\frac{a}{8M^{2}}\right)^{2}r_{G}^{2}\sin^{2}{\theta}}}.
\end{align}

In the pulsar case, we defined the ``radial velocity'' such that outgoing light rays have unit coordinate speed, allowing for a physically transparent visualization of an outgoing flow.  In the black hole case we have both outgoing and ingoing flows, so the analogous approach requires two separate definitions.  For the outgoing flow we want to use $\tilde{t}_{\rm out}$ such that $dr/d\tilde{t}_{\rm out}=1$ for outgoing null rays, and for the ingoing flow we want to use $\tilde{t}_{\rm in}$ such that $dr/d\tilde{t}_{\rm in}=-1$ for ingoing null rays.  The time coordinates of the ingoing/outgoing Kerr-Schild coordinates (Appendix \ref{app: Kerr-Schild coords}) satisfy these properties. Since we work to the first order in $a$, those coordinates are expanded in $a$ consistently. 

Fig.~\ref{fig: BZ velocity} shows the radial velocities $dr/d\tilde{t}_{\rm in/out}$ on the field sheet $\theta=\pi/6$. (Above the gap, $\tilde{t}_{\rm out}$ is used, while below the gap, $\tilde{t}_{\rm in}$ is used.) We chose the gap location to be $r_{G}=3M$ and set $a/M=0.1$, then vary the initial conditions.  As in the pulsar case, we obtain non-MHD behavior for low multiplicity, with parallel electric field (Fig.~\ref{fig: BZ E0}) supporting the differential motion of electrons and positrons.

\section{Structure of the perturbation expansion}\label{sec:consistency}

In contrast to a typical perturbation expansion where all quantities at a given order are determined simultaneously, the expansion we consider has an intricate staggered structure, with the various quantities determined in a specific sequence of operations.  So far, we followed this sequence long enough to determine the leading parallel corrections to FFE, i.e., the plasma parallel motion and parallel electric field.  In this section we explore how the chain of logic could be continued to determine perpendicular motion (drift) as well as higher-order corrections to all quantities.

Let us first restate the procedure used so far. 
To zeroth-order in $\epsilon$, we have Eqs.~(\ref{eq:EoM 0th})--(\ref{eq:Maxwell2 0th}). 
As explained, these conditions immediately imply that the zeroth order field strength is degenerate, so that we have well-defined parallel and perpendicular planes at each point. The parallel electric field and perpendicular velocity are automatically vanishing by these same equations.  The background fluid species obey the Lorentz force law with the perturbed Maxwell field \eqref{eq:EoM 1st}, an order-mixing that arises because the perturbation parameter appears in the equations.  After projecting this law onto the parallel plane, only the parallel portion of the perturbed Maxwell field appears, and this portion can be eliminated \eqref{eq: consistency} by adding together the equations for the positive and negative species (since both species feel the same electric field).  Combined with the the continuity equations \eqref{eq:continuity 0th} and the condition that the plasma match the force-free current \eqref{eq: current=plasma flow 0th}, we obtain coupled equations for $n_{\pm}$ and $u^a$ depending only on the force-free fields, together with an explicit formula \eqref{eq: u and E0} for the perturbed parallel electric field $\delta E_0$.

We may continue this procedure by projecting all equations onto the parallel and perpendicular planes.  Perturbing the field equations 
\eqref{eq:Maxwell1} with \eqref{eq: current=plasma flow} and projecting to the perpendicular plane via contraction with $F\indices{^{a}_{b}}$ gives
\begin{align}
    \label{eq:maxwell1 1st}
    &F\indices{^{a}_{b}} \nabla_{c}\delta F^{bc}\notag
    \\
    &=  {F}\indices{^{a}_{b}} 
    \left(e\delta n_{+}u_{+}^{b}+en_{+}\delta u_{+}^{b}
     -e\delta n_{-}u_{-}^{b}-en_{-}\delta u_{-}^{b}\right)\notag
     \\
     &=en_{+} F^{ab}\delta u_{+b}-en_{-} F^{ab}\delta u_{-b}\notag
     \\
     &=en_{+}\left(+u_{+}^{c}\nabla_{c}u_{+}^{a}-\delta F^{ab}u_{+b}\right)\notag
     \\
     &\hspace{2cm}-en_{-}\left(-u_{-}^{c}\nabla_{c}u_{-}^{a}-\delta F^{ab}u_{-b}\right)\notag
     \\
     &=
     \left(-en_{+}u_{+b}+en_{-}u_{-b}\right)\delta F^{ab}\notag
     \\
     &
     \hspace{2cm}+en_{+}u_{+}^{c}\nabla_{c}u_{+}^{a}
     +en_{-}u_{-}^{c}\nabla_{c}u_{-}^{a},
\end{align}
where Eqs.~(\ref{eq:EoM 0th}) and (\ref{eq:EoM 1st}) have been used.  This eliminates the first-order velocity, leaving the definite equation \eqref{eq:maxwell1 1st} for the field perturbation $\delta F$.  Thus, together with the magnetic flux conservation 
\begin{align}
    \label{eq:maxwell2 1st}
    \nabla_{b}\delta \tilde{F}^{ab}=0,
\end{align} 
we have a linear equation for $\delta F^{ab}$ that does not involve $\delta u^a_\pm$.  Solving this equation would determine the first-order field without reference to the first-order plasma.  This is analogous to FFE at zeroth order.

Next we may determine the perpendicular plasma motion. Recall that the zeroth-order parallel motion was determined by projecting Eq.~(\ref{eq:EoM 1st}) onto the field sheet.  The first-order perpendicular motion instead comes from projecting onto the perpendicular plane.  
Using Eq.~(\ref{eq: F=B epsilon perp}), one finds
\begin{align}              
    \epsilon_{\perp ab}\epsilon^{db}_{\perp}\delta u_{\pm b}
    &= \epsilon_{\perp ad}\left(\pm\frac{1}{B_{0}}
    u^{c}_{\pm}\nabla_{c}u^{d}_{\pm}
    +\frac{1}{B_{0}}\delta F^{db}u_{\pm b}\right),\notag
    \end{align}
which via (\ref{eq: hperp and epsilonperp}) means that the perpendicular component $\delta u_{\perp a}\equiv{h_{\perp ab}}\delta u^{b}$ is given by
    \begin{align}
    \label{eq: delta u perp}
    \delta u_{\pm\perp a}
    &=\pm \frac{1}{B_{0}} \epsilon_{\perp ad}
    u^{c}_{\pm}\nabla_{c}u^{d}_{\pm}
    +\frac{1}{B_{0}} \epsilon_{\perp ad}\delta F^{db} u_{\pm b}.
\end{align}
These terms coincide with the traditional drift velocities (for example, the zero-gyration limit of Eq.~(222) in \citep{Vandervoort1960jul}).  Using \eqref{eq: delta u perp}, the perpendicular velocity is immediately determined from the background flow and first-order field.

The parallel velocity $\delta u_{\pm\parallel}^{a}$ needs to be determined in a similar way to how we determined the zeroth-order parallel velocity using the consistency condition (\ref{eq: consistency}): we write down the second-order equation of motion and the condition that both species have to feel the same electric field provides a constraint for the parallel velocity.  
It is easily verified that the second-order equation of motion for the plasma is given as follows:
\begin{align}
    \label{eq:2nd EoM}
    &\pm\left(
    \delta u_{\pm}^{b}\nabla_{b}u_{\pm}^{a}
    +
    u_{\pm}^{b}\partial_{b}\delta u_{\pm}^{a}
    \right)\notag
    \\
    &\quad=
    F^{ab}\delta u_{\pm b}^{(2)}
    +\delta F^{ab}\delta u_{\pm b}
    +\delta F^{(2)ab}u_{\pm b},
\end{align}
where $\delta u_{\pm}^{(2)a}$ and $\delta F^{(2)}_{ab}$ denote the second-order quantities. 
Now, we consider multiplying $u_{\pm}^{c}\tilde{F}^{}_{ca}$ with this expression (which is equivalent to projecting Eq.~(\ref{eq:2nd EoM}) onto the field sheet and extracting the component of Eq.~(\ref{eq:2nd EoM}) in the parallel direction to $u_{\pm}^{a}$). 
Then, the first term on the RHS of Eq.~(\ref{eq:2nd EoM}) becomes
\begin{align}
    u_{\pm}^{c}\tilde{F}^{}_{cb}
     F^{ba}\delta u_{\pm a}^{(2)}
     =0
\end{align}
due to the force-free condition (\ref{FFE}). 
The second term on the RHS of Eq.~(\ref{eq:2nd EoM}) can be simplified by using Eq.~(\ref{eq:1 species EoM 1st}) and noticing that $u_{\pm}^{c} \delta u_{\pm c}=0$ is satisfied:
\begin{align}
     &u_{\pm}^{c}\tilde{F}_{cb}
     \delta F^{ba}\delta u_{\pm a}\notag
     \\
     &=
       u_{\pm}^{c}
       \left[
       -\delta\tilde{F}_{cb}
     F^{ba}
     -\frac{1}{2}
     \delta^{a}_{c}
     \delta\tilde{F}_{ef}
     F^{ef}
       \right]
     \delta u_{\pm a}\notag
     \\
     &=
     - u_{\pm}^{c}\delta\tilde{F}_{cb}
     \left[
     - \delta F^{ba} u_{\pm a}
     \pm
     u_{\pm}^{\rho}\nabla_{\rho}u_{\pm}^{b}
     \right]\notag
     \\
     &=
      u_{\pm}^{c}\delta\tilde{F}_{cb} \delta F^{ba} u_{\pm a}
      \mp 
      u_{\pm}^{c}\delta \tilde{F}_{cb}u_{\pm}^{d}\nabla_{d}u_{\pm}^{b}.
\end{align}
The first term of the above expression can be further simplified as follows:
\begin{align}
     u_{\pm}^{c}\delta\tilde{F}_{cb} \delta F^{ba} u_{\pm a}
     &=
      u_{\pm}^{c}
      \left[
      -\delta\tilde{F}_{cb} \delta F^{ba}
      -\frac{1}{2}
      \delta^{a}_{c}
      \delta\tilde{F}_{ef} \delta F^{ef}
      \right]
      u_{\pm a}.
\end{align}
Since the LHS and the first term on the RHS are identical, solving for this leads to 
\begin{align}
     u_{\pm}^{c}\delta\tilde{F}_{cb} \delta F^{ba} u_{\pm a}
       &=\frac{1}{4}\delta\tilde{F}_{ef} \delta F^{ef}.
\end{align}
Thus, the projection of the second term on the RHS of Eq.~(\ref{eq:2nd EoM}) is 
\begin{align}
    u_{\pm}^{c}\tilde{F}_{cb}
     \delta F^{ba}\delta u_{\pm a}&=
     \frac{1}{4}\delta\tilde{F}_{ef} \delta F^{ef}
     \mp 
      u_{\pm}^{c}\delta\tilde{F}_{c b}u_{\pm}^{d}\nabla_{d}u_{\pm}^{b}.
\end{align}

The last term can be simplified in a similar manner:
\begin{align}
    &u_{\pm}^{c}\tilde{F}_{cb}
     \delta F^{(2)ba}u_{\pm a}\notag
     \\
     &=
       u_{\pm}^{c}
       \left[
       -\delta\tilde{F}^{(2)}_{cb}
     F^{ba}
     -\frac{1}{2}
     \delta^{a}_{c}
     \delta \tilde{F}^{(2)}_{ef}
     F^{ef}
       \right]
     u_{\pm a}\notag
     \\
     &=\frac{1}{2}\delta\tilde{F}^{(2)}_{ef}
     F^{ef}
\end{align}

In total we have 
\begin{align}\label{eq:2nd EoM 2}
     &\pm u_{\pm}^{c}\tilde{F}_{cb}\left(
    \delta u_{\pm}^{d}\nabla_{d}u_{\pm}^{b}
    +
    u_{\pm}^{d}\nabla_{d}\delta u_{\pm}^{b}
    \right)      \pm   u_{\pm}^{c}\delta\tilde{F}_{cb}u_{\pm}^{d}\nabla_{d}u_{\pm}^{b}\notag
    \\
    &\quad=
    \frac{1}{2}\delta \tilde{F}^{(2)}_{ef}
     F^{ef}
     + 
     \frac{1}{4}\delta\tilde{F}_{ef} \delta F^{ef}\notag
     \\
     &\quad=\left(\delta E_{0}\right)^{2}+\delta E_{0}^{(2)}B_{0}
\end{align}
which is analogous to Eq.~(\ref{eq:1 species 2d EoM}).  The unknown second-order perturbation $\delta E_{0}^{(2)}$ can be analogously eliminated by subtracting the equations for the positive and negative species, yielding
\begin{align}
        \label{eq:alphae alphap}
         &u_{+}^{c}\tilde{F}_{cb}(
    \delta u_{+}^{d}\nabla_{d}u_{+}^{b}\notag
   +
     u_{+}^{d}\nabla_{d}\delta u_{+}^{b})
     +
      u_{+}^{c}\delta\tilde{F}_{cb}
      u_{+}^{d}\nabla_{d}u_{+}^{b}
      \\
      +&
       u_{-}^{c}\tilde{F}_{cb}(
    \delta u_{-}^{d}\nabla_{d}u_{-}^{b}
    +
     u_{-}^{d}\nabla_{d}\delta u_{-}^{b})
     +
      u_{-}^{c}\delta\tilde{F}_{cb}
      u_{-}^{d}\nabla_{d}u_{-}^{b}
      =0,
\end{align}
which is analogous to Eq.~\eqref{eq: consistency}.  The only unknowns in this equation are the velocity perturbations $\delta u^a_\pm$.  These perturbations also obey the continuity equation,
\begin{align}\label{eq:1st conservation}
    \nabla_{a}\left(\delta n_{\pm}u_{\pm}^{a}+n_{\pm}\delta u_{\pm}^{a}\right)=0.
\end{align}
In addition, the plasma flow has to produce the already known current according to Eq.~(\ref{eq: current=plasma flow}), and to first order, it is given by 
\begin{align}
    \label{eq: current=plasma flow 1st}
    \delta j^{a}=e \left(n_{+}\delta u_{+}^{a}+\delta n_{+}u^{a}_{+}-n_{-}\delta u_{-}^{a}-\delta n_{-}u^{a}_{-}\right).
\end{align} 
Eq.~\eqref{eq: delta u perp} has already provided the perpendicular velocity perturbations, and Eqs.~\eqref{eq:alphae alphap}--(\ref{eq: current=plasma flow 1st}) constitute three equations for the parallel velocity perturbations and the density perturbations $\delta n_\pm$. 

Once the plasma density and velocity perturbations are determined by solving these equations, we can determine the second-order parallel electric field perturbation from a simple rearrangement of Eq.~\eqref{eq:2nd EoM 2},
\begin{align}\label{eq:2nd E0}
     \delta E_{0}^{(2)}=
     &B_0^{-1}\Big(\pm u_{\pm}^{c}\tilde{F}_{cb}\left(
    \delta u_{\pm}^{d}\nabla_{d}u_{\pm}^{b}
    +
    u_{\pm}^{d}\nabla_{d}\delta u_{\pm}^{b}
    \right)     \notag \\ & \pm   u_{\pm}^{c}\delta\tilde{F}_{cb}u_{\pm}^{d}\nabla_{d}u_{\pm}^{b} - \left(\delta E_{0}\right)^{2}\Big).
\end{align}
This equation is analogous to Eq.~\eqref{eq: u and E0} and brings us one order higher in perturbation theory than considered in the main text.  Higher orders would follow the same pattern, as summarized in Table.~\ref{fig: perturbation procedure}.

\renewcommand{\arraystretch}{1.2}

\definecolor{gry}{gray}{0.7}
\definecolor{buff}{rgb}{0.94, 0.86, 0.51}

\begin{table*}
\centering
\resizebox{15cm}{!}{%
\begin{NiceTabular}{ccccc}[hvlines]
\CodeBefore
\Body
    \diagbox{}{} & \makecell{$E_{0}$ \\ (explicit)}    & \makecell{$F_{\mu\nu}$\\(PDE)} &\makecell{$u_{\perp}$\\(explicit)} &\makecell{$n_{\pm}$  $u_{\parallel}$\\(PDE)}
  \\
  0th    
  &0
  &FFE
  &0
  &
  Eqs.~\eqref{eq:continuity 0th}, \eqref{eq: current=plasma flow 0th}, \eqref{eq: consistency}
  \\
  & \multicolumn{4}{c}{\phantom{a}} 
  \\
  1st     
  &\phantom{a}$\delta E_{0}=\mp \epsilon_{ab}D^{a}u^{b}_{\pm}$\phantom{a}
  &\phantom{a}Eqs.~(\ref{eq:maxwell1 1st}) (\ref{eq:maxwell2 1st})\phantom{a}
  &\phantom{a}\makecell{drift vel.\\Eq.~(\ref{eq: delta u perp})}\phantom{a}
  &\phantom{a}Eqs.~(\ref{eq:alphae alphap}) (\ref{eq:1st conservation}) (\ref{eq: current=plasma flow 1st})\phantom{a}
  \\
  & \multicolumn{4}{c}{\phantom{a}}
  \\
  2nd  
  &Eq.~(\ref{eq:2nd E0})
  &\phantom{a}
  &\phantom{a}
  &\phantom{a}
  \\
  & \multicolumn{4}{c}{\phantom{a}} 
  \\
  3rd 
  &\phantom{a}
  &\phantom{a}
  &\phantom{a}
  &\phantom{a}
  \\
  & \multicolumn{4}{c}{$\vdots$}
  \\
  \CodeAfter
  \begin{tikzpicture} [shorten < = 0.0mm, shorten > = 0.0mm, black, ->,line width=0.2mm] 
  \draw([xshift=8mm,yshift=0mm]2-2.center) -- ([xshift=-8mm,yshift=0mm]2-3.center);
  \draw([xshift=12mm,yshift=0mm]4-2.center) to ([xshift=-12mm,yshift=0mm]4-3.center); 
  \draw([xshift=8mm,yshift=0mm]6-2.center) to ([xshift=0mm, yshift=0mm]6-3.center); 
  \draw([xshift=0mm,yshift=0mm]8-2.center) to ([xshift=0mm, yshift=0mm]8-3.center); 
  \end{tikzpicture}
  \begin{tikzpicture} [shorten < = 0.0mm, shorten > = 0.0mm, black, ->,line width=0.2mm]
  \draw ([xshift=8mm,yshift=0mm]2-3.center) -- ([xshift=-5mm,yshift=0mm]2-4.center) ; 
  \draw ([xshift=12mm,yshift=0mm]4-3.center) -- ([xshift=-7mm,yshift=0mm]4-4.center) ; 
  \draw (6-3) -- (6-4) ; 
  \draw (8-3) -- (8-4) ; 
  \end{tikzpicture}
   \begin{tikzpicture} [shorten < = 0.0mm, shorten > = 0.0mm, black, ->,line width=0.2mm] 
  \draw([xshift=5mm,yshift=0mm]2-4.center) to ([xshift=-15mm, yshift=0.1mm]2-5.center);
  \draw([xshift=8mm,yshift=0mm]4-4.center) to ([xshift=-15.5mm, yshift=0mm]4-5.center); 
  \draw([xshift=0mm,yshift=0mm]6-4.center) to ([xshift=0mm, yshift=0mm]6-5.center); 
  \draw([xshift=0mm,yshift=0mm]8-4.center) to ([xshift=0mm, yshift=0mm]8-5.center); 
  \end{tikzpicture}
  \begin{tikzpicture} [shorten < = 0.0mm, shorten > = 0.2mm, black, ->,line width=0.2mm]
  \draw ([yshift=-2mm]2-5.center) -- ([yshift=2mm]4-2.center) ; 
  \draw ([yshift=-3mm]4-5.center) -- ([yshift=3mm]6-2.center) ; 
  \draw ([yshift=-1mm]6-5.center) -- ([yshift=1mm]8-2.center) ; 
  \end{tikzpicture}
\end{NiceTabular}
}
\caption{\justifying Structure of the perturbation procedure.  Rows represent orders in $\epsilon=m/e$, while columns indicate quantities to be determined.  The logical flow is represented by arrows, such that by following the arrows, each quantity depends only on quantities previously determined.  The parallel electric field and perpendicular plasma flow are given by explicit equations and therefore uniquely and immediately determined without additional boundary conditions.  By contrast, the full electric field and full plasma require solving partial differential equations with suitable boundary conditions.} 
\label{fig: perturbation procedure}
\end{table*}

\section{Outlook}\label{sec:outlook}

In this paper we have developed a general method for determining the behavior of two-fluid plasma in the nearly force-free regime.  We derived general equations and showed how they can be solved algebraically when there is a symmetry adapted to the magnetic field configuration.  We demonstrated the method in the case of pulsar and black hole magnetospheres with radial field lines.

There are several natural next steps.  First, one could move beyond radial field lines to consider magnetospheres with dipolar or other magnetic field lines.  This would require a numerical simulation to determine the background force-free configuration, after which the plasma could be added by solving algebraic equations built from the numerical results.  This could be done both for aligned and inclined magnetospheres, since the latter still possesses enough symmetry to reduce the plasma equations to algebraic equations.  In this way, the plasma flow could be determined as a function of gap boundary conditions in more realistic magnetospheric configurations.

Another path forward is to move beyond the two-fluid approximation.  As outlined in the introduction, any valid description of plasma should reproduce FFE in the strong-magnetic-field limit, and should therefore have a staggered perturbation expansion analogous to the one we study in the two-fluid case.  It would be interesting to develop this approximation for kinetic theory or other descriptions of the plasma microphysics.  

We hope that the results of this paper will form a foundation for future perturbative investigations of magnetically dominated plasma.

\section*{Acknowledgements}
This work was supported by grants from the Simons Foundation (MP-SCMPS-00001470) and the National Science Foundation (PHY-230919, PHY-2513082). AP additionally acknowledges support by Alfred P.~Sloan Research Fellowship and a Packard Foundation Fellowship in Science and Engineering.

\appendix

\section{A quick review of FFE}\label{sec:FFE}

Consider an electromagnetic field $F_{ab}$ propagating in a (possibly curved) spacetime $g_{ab}$ along with some charged matter.  Irrespective of how the matter behaves, the field must obey the homogeneous Maxwell equations
\begin{align}
    \nabla_a \tilde{F}^{ab} = 0,
\end{align} 
which  can be thought of as conservation of magnetic flux.  If charged matter is energetically negligible, then the field also must obey conservation of energy and momentum,
\begin{align}
    \nabla_a T^{ab}_{\rm EM} = 0.
\end{align}
This equation may equivalently be written
\begin{align}\label{Fdotj}
    F_{ab} j^b = 0,
\end{align}
where $j^a = \nabla_{b} F^{ab}$ is the charge-current that would be needed to source this field strength.  This expresses the vanishing of the Lorentz force density and is called the force-free condition.

The validity of the force-free condition requires negligible exchange of energy and momentum with charged matter.  If the charged matter is \textit{entirely} negligible, $j^a = 0$, then we recover the vacuum Maxwell equations.  However, if we insist that the charged matter does still influence the electromagnetic field, i.e. $j^a \neq 0$, then a very interesting alternative theory emerges.  Equation \eqref{Fdotj} now shows that the field strength is degenerate, $F_{ab} \tilde{F}^{ab}=0$ (equivalently $\vec{E}\cdot \vec{B}=0$), which can be used to express \eqref{Fdotj} in an explicit evolution form, which turns out to be deterministic with the additional assumption that $F_{ab} F^{ab}>0$ (equivalently $B^2>E^2$) \citep{Palenzuela.Bona.ea2011jun,Pfeiffer.MacFadyen2013jul,Carrasco.Reula2016apr}.  These equations together are known as force-free electrodynamics, listed in Eq.~\eqref{FFE} of the main body.

The degeneracy and magnetic domination of the Maxwell field together have the physical interpretation that the plasma is fully screened---free charges always rearrange to eliminate the electric field.  While we assumed non-zero charge-current ($j^a \neq 0$) to derive the screening, nothing precludes a fully screened plasma from having $j^a=0$, entering a state of zero \textit{net} charge and current.  We therefore find it more natural to drop the assumption $j^a \neq 0$ and instead regard algebraic conditions on the field strength ($F_{ab} \tilde{F}^{ab}=0$ and $F_{ab}F^{ab}>0$) as fundamental.  That is, we take  Eqs.~\eqref{FFE} to be our definition of a force-free field, without any additional restriction on the charge-current.  This allows vacuum Maxwell solutions to count as ``force-free'' provided they are degenerate and magnetically dominated.  Such solutions arise frequently when approximation methods are employed to solve the force-free equations \cite{Blandford.Znajek1977jul,Gralla.Zimmerman2016jun,Gralla.Lupsasca.ea2016dec,Gralla.Lupsasca.ea2017dec,Armas.Cai.ea2020apr}.

\section{Field eigenframes}\label{app: field eigenframe}

Let $a$ and $b$ denote the natural Lorentz scalars associated with an electromagnetic field,
\begin{align}
    \label{eq:invariant a}
    a & =\frac{1}{2}F^{ab}F_{ab }=\vec{B}^2-\vec{E}^2, \\
    \label{eq:invariant b}
    b & = \frac{1}{2}F^{ab}\tilde{F}_{ab} = 2 \vec{E}\cdot\vec{B}.
\end{align}
We can relate these to invariant electric and magnetic scalars $E_0$ and $B_0$ via
\begin{align}\label{BEab}
    B_{0}^{2}-E_{0}^{2}= a, \qquad   2E_{0}B_{0} = b.
\end{align}
For a magnetically dominated field, we have $|B_0|>|E_0|$ and hence $B_0 \neq 0$.  We choose $B_{0}>0$ without loss of generality and solve \eqref{BEab} as
\begin{align}
    \sqrt{2}E_{0}&=\mathrm{sign}(b)\sqrt{\sqrt{a^{2}+b^{2}}-a},
    \\
    \sqrt{2}B_{0}&=\sqrt{\sqrt{a^{2}+b^{2}}+a}.
\end{align}
Given the assumption of magnetic domination, it is always possible to boost into a frame where $\vec{E}$ and $\vec{B}$ are parallel (including the special case of degenerate fields, where $\vec{E}=0$ in such a frame).  We may further rotate to make $z$ the magnetic field direction, so that $\vec{B}=(0,0,B_0)$ and $\vec{E}=(0,0,E_0)$.  We may think of the $tz$ subspace as the timelike ``field plane'' and the $xy$ subspace as the spacelike ``perpendicular plane''.  We will denote the projection tensors to these planes by $h_{ab}$ and $(h_{\perp})_{ab}$, respectively.  Adopting the signs $\epsilon_{tz}=-1$ and $(\epsilon_\perp)_{xy}=+1$ for the antisymmetric projection tensors, covariant expressions for these quantities are
\begin{align}
    h^{ab}&=\frac{F\indices{^a_c}F^{cb}+B_{0}^{2}g^{ab}}{E_{0}^{2}+B_{0}^{2}}
    =\frac{\tilde{F}\indices{^a_c}\tilde{F}^{cb}+E_{0}^{2}g^{ab}}{E_{0}^{2}+B_{0}^{2}},
    \\
     h_{\perp}^{ab}&=\frac{-F\indices{^a_c}F^{cb}+E_{0}^{2}g^{ab}}{E_{0}^{2}+B_{0}^{2}}
    =\frac{-\tilde{F}\indices{^a_c}\tilde{F}^{cb}+B_{0}^{2}g^{ab}}{E_{0}^{2}+B_{0}^{2}},
    \\
     \epsilon^{ab}&=-\frac{F\indices{^{a}_{c}}h^{cb}}{E_{0}}
     =\frac{\tilde{F}\indices{^{a}_{c}}h^{cb}}{B_{0}}
     =\frac{B_{0}\tilde{F}^{ab}-E_{0}F^{ab}}{E_{0}^{2}+B_{0}^{2}},
     \\
     \epsilon_{\perp}^{ab}&=\frac{F\indices{^{a}_{c}}h_{\perp}^{cb}}{E_{0}}
     =\frac{\tilde{F}\indices{^{a}_{c}}h_{\perp}^{cb}}{B_{0}}
     =\frac{E_{0}\tilde{F}^{ab}+B_{0}F^{ab}}{E_{0}^{2}+B_{0}^{2}}.
\end{align}
The symmetric tensors decompose the metric as
\begin{align}
    g^{ab}=h^{ab}+h_{\perp}^{ab}
   =- \frac{F\indices{^a_c}\tilde{F}^{cb}}{E_{0}B_{0}},
\end{align}
while the antisymmetric tensors decompose the field strength and dual as
\begin{align}
    \label{eq:F decompsition}
    F^{ab}&=-E_{0}\epsilon^{ab}+B_{0}\epsilon_{\perp}^{ab},
    \\
    \label{eq:F dual decompsition}
    \tilde{F}^{ab}&=B_{0}\epsilon^{ab}+E_{0}\epsilon_{\perp}^{ab}.
\end{align}

Some other useful relations include the action of the dual operator,
\begin{align}
    \epsilon_{\perp ab}&=-*\epsilon_{ab},
    \\
    \epsilon_{ab} &= *\epsilon_{\perp ab},
\end{align}
normalization relations
\begin{align}
    \epsilon^{ab}\epsilon_{ab}&=-2,
    \\
    \epsilon_{\perp}^{ab}\epsilon_{\perp ab} & = 2,
    \\ 
    \epsilon^{ac}\epsilon_{\perp cb}&=0,
\end{align}
as well as
\begin{align}
    h^{ab}={\epsilon}\indices{^a_{c}}{\epsilon}\indices{^{cb}},\quad
    h_{\perp}^{ab}=-{\epsilon_{\perp}}\indices{^a_{c}}{\epsilon_{\perp}}\indices{^{cb}}.
\end{align}

The field plane is timelike and hence has two null directions satisfying $h_{ab} \ell^b=\ell_a$.  Taking $\ell_+$ to move in the magnetic field direction (positive $z$ in the special coordinates) the two null directions $\ell_{\pm}$ satisfy 
\begin{align}
     \label{eq:h lpm= lpm}
     h^{ab}\ell_{\pm b}&=\ell_{\pm}^{a}, 
     \\
     \label{eq:epsilon lpm=mp lpm}
     \epsilon^{ab}\ell_{\pm b}&=\mp \ell_{\pm}^{a}.
\end{align}
We may then plug into \eqref{eq:F decompsition} and \eqref{eq:F dual decompsition} to find
\begin{align}
    F\indices{^{a}_{b}}\ell_{\pm}^{b}& = \pm E_{0}\ell_{\pm}^{a} \\
    \label{eq:Ftilde PND =B0 PND}
    \tilde{F}\indices{^{a}_{b}}\ell_{\pm}^{b}& = \mp B_{0}\ell^{\pm a}.
\end{align}
We see that $\ell_\pm^a$ are the eigenvectors of the field strength with eigenvalue $\pm E_0$, and also the eigenvectors of the dual field strength with eigenvalue $\mp B_0$.  These are known as the principal null directions of the field.  In the text we denote them $\ell^a=\ell^a_+$ and $n^a=\ell^a_-$ to avoid confusion with the $\pm$ sign that describes the sign of the charge.  

For degenerate fields ($E_0=0$), both $\ell_+$ and $\ell_-$ are degenerate as eigenvectors of the field strength, having the same (zero) eigenvalue.  However, they are non-degenerate as eigenvectors of the dual field strength, having distinct eigenvalues $\mp B_0$.

\section{Estimation of the electric field near the light cylinder}\label{app: strong E at LC}
In Fig.~\ref{fig: Michel r velocity e p} we demonstrated the phenomenon of light cylinder acceleration when the initial positron velocity is very small (Lorentz factor $\gamma^*_+\sim 1$).  In this appendix we compute an analytical approximation for this behavior, treating $\gamma^+_*-1$ as a small parameter,
\begin{align}
    \Delta \gamma_+^* \equiv \gamma_+^*-1 \ll 1.
\end{align}
First we will show that small $\Delta \gamma^*_+$ is the same approximation as small $\mu_+$.  In the small $\Delta \gamma^*_{+}$ limit, we can find the expression for $\beta_{+}$ at $R_{*}$ by solving Eq.~(\ref{eq:boundary gammap}) as
\begin{align}
    \label{eq:beta- surface}
    \beta_{+}|_{R_{*}}=\sqrt{-\chi^{a}\chi_{a}|_{R_{*}}}\left(1+\sqrt{2\Delta \gamma_{+}^*}\right)+\mathcal{O}\left(\Delta \gamma_{+}^*\right).
\end{align}
For Eq.~(\ref{eq:boundary lambda}), by using Eqs.~(\ref{eq: current=plasma flow 0th}), (\ref{eq: define mu}), (\ref{eq: chi^2=2chiell}), (\ref{eq: mono jchi}), we find
\begin{align}
    \lambda_{+}^*
    &=-\left.\frac{en_{+}u_{+}^{a}\chi_{a}}{|j^{a}\chi_{a}|}\right|_{R_{*}}\notag
    \\
    &=\mu_{+} \left.\frac{\epsilon_{ab}j^{a}\chi^{b}}{j^{a}\chi_{a}}\frac{u_{+}^{a}\chi_{a}}{\epsilon u_{+}^{a}\chi^{b}}\right|_{R_{*}}\notag
    \\
    &=
    \left.\mu_{+}\frac{\beta_{+}\ell_{+}^{a}\chi_{a}+\frac{1}{\beta_{+}}\ell_{-}^{a}\chi_{a}}{\beta_{+}\ell_{+}^{a}\chi_{a}-\frac{1}{\beta_{+}}\ell_{-}^{a}\chi_{a}}\right|_{R_{*}}\notag
    \\
    &=
    \left.\mu_{+}\frac{\beta_{+}^{2}-\chi^{a}\chi_{a}}{\beta_{+}^{2}+\chi^{a}\chi_{a}}\right|_{R_{*}}. \label{lambdaplusstar}
\end{align}
We have also used $\epsilon^{ab}\ell_b=-\ell^a$, which is just Eq.~(\ref{eq:epsilon lpm=mp lpm}) in the notation of the main body.  To get rid of the absolute value, we have used the fact that $j^{a}\chi_{a}>0$ for a region $\theta<\pi/2$ as we can see in Eq.~(\ref{eq: mono jchi}).  From Eqs.~\eqref{eq:beta- surface} and \eqref{lambdaplusstar}, we see that $\mu_{+}$ is perturbatively given by 
\begin{align}
    \label{eq:mu+ small gammap}
    \mu_{+}=\lambda_{+}^*\sqrt{2\Delta \gamma_{+}^*}+\mathcal{O}\left(\Delta\gamma_{+}^*\right).
\end{align}
This implies that the small $\Delta \gamma_{+}^*$ limit corresponds to the small $\mu_{+}$ limit.  We will consider the small $\mu_{+}$ limit instead of the small $\Delta \gamma_{+}^*$ limit since the fact that $\mu_{+}$ is constant in the entire magnetosphere makes it easier to take the desired limit. 

Our strategy for estimating the maximum value of $\delta E_{0}$ is based on the observation (in our numerical studies) that just outside the light cylinder, the particle velocity rapidly transitions between a nearly flat profile across the light cylinder and its eventual asymptotic profile at large $r$.  We solve for each profile separately (in terms of conserved quantities) and set them equal to obtain an estimate of the transition radius.  This rough estimate turns out to provide the radius of maximum $\delta E_{0}$ with good accuracy (see Fig.~\ref{fig: Michel E0 SA}).  A more accurate estimate would involve constructing a special limit that zooms in on the transition region (just outside the light cylinder) and solves the equations in detail.  Such an estimate is not necessary for our purposes.

With this strategy in mind, let us first evaluate $\beta_{\pm}$ at the LC.  At the LC, the corotating frame becomes null, i.e. $\chi^{a}\chi_{a}=0$. 
Due to Eq.~(\ref{eq: chi^2=2chiell}), this condition is equivalent to 
\begin{align}
    \label{}
    n^{a}\chi_{a}|_{\rm LC}=0.
\end{align}
Using this result together with Eq.~(\ref{eq: cnsrv2}), we can simplify the expression for $n_{\pm}$ at the LC as
\begin{align}
    n_{\pm}|_{\mathrm{LC}}=\left.-2\frac{C_{\pm}B_{0}}{\beta_{\pm}}\right|_{\mathrm{LC}}.
\end{align}
In addition, by dotting Eq.~(\ref{eq: current=plasma flow 0th}) with $\ell$  and using the fact that $j^{a}$ is proportional to $\ell$, we have
\begin{align}
    \frac{n_{+}}{\beta_{+}}=\frac{n_{-}}{\beta_{-}}.
\end{align}
Therefore, at the LC, we obtain the following relation between $\beta_{+}$ and $\beta_{-}$
\begin{align}
    \frac{\mu_{+}}{{\beta_{+}^{2}}|_{\mathrm{LC}}}=\frac{\mu_{-}}{{\beta_{-}^{2}}|_{\mathrm{LC}}}.
\end{align}
Furthermore, evaluating Eq.~(\ref{eq: cnsrv1}) at the LC, we find
\begin{align}
    \beta|_{+\mathrm{LC}}+\beta|_{-\mathrm{LC}}=2\Gamma.
\end{align}
In the small $\mu_{+}$ limit of interest to light cylinder acceleration, $\beta_{\pm\mathrm{LC}}$ are found to be
\begin{align}
    \label{eq:beta+ near LC}
    \beta_{+}|_{\mathrm{LC}}&=2\Gamma\sqrt{\mu_{+}}+\mathcal{O}(\mu_{+}),
    \\
     \label{eq:beta- near LC}
    \beta_{-}|_{\mathrm{LC}}&=2\Gamma+\mathcal{O}\left(\sqrt{\mu_{+}}\right).
\end{align}

Next, we consider the behavior of $\beta_{\pm}$ in the large $r$ region. 
In such a region, the value of $\chi^{a}\chi_{a}$ becomes large and from Eq.~(\ref{algebraic eq2}), it is implied that $\beta_{\pm}$ must scale as
\begin{align}
\beta_{\pm}\sim\sqrt{n^{a}\chi_{a}}\sim\sqrt{\chi^{a}\chi_{a}}.
\end{align}
To capture how $\beta_{\pm}$ behaves in this region, we introduce a fiducial expansion parameter $\delta$ as $\chi^{a}\chi_{a}\to \chi^{a}\chi_{a}/\delta^{2} $. 
Using $\delta$, we can consider the large $r$ limit by taking the small $\delta$ limit. 
To compute $\beta_{\pm}$, we can expand $\beta_{\pm}$ in the power of $\delta$ and, then by taking $\delta=1$ at the end, we can obtain the perturbative approximation to $\beta_{\pm}$. 
In other words, we can write Eqs.~(\ref{algebraic eq1}),(\ref{algebraic eq2}) as
\begin{align}
  2\Gamma&=\left(\beta_{+}+\beta_{-}\right)
    -\left(\frac{1}{\beta_{+}}+\frac{1}{\beta_{-}}\right)\frac{\chi^{a}\chi_{a}}{\delta^{2}},
    \\
     -1
    &=\mu_{+}\frac{\beta_{+}^{2}-(\chi^{a}\chi_{a}/\delta^{2})}{\beta_{+}^{2}+(\chi^{a}\chi_{a}/\delta^{2})}
    -
    \mu_{-}\frac{\beta_{-}^{2}-(\chi^{a}\chi_{a}/\delta^{2})}{\beta_{-}^{2}+(\chi^{a}\chi_{a}/\delta^{2})},
\end{align}
and can solve these equations perturbatively in $\delta$ with the expansion ansatz for $\beta_{\pm}$ given as
\begin{align}
    \beta_{+}&=\frac{1}{\delta} \beta_{+}^{(-1)}+\beta_{+}^{(0)}+\mathcal{O}(\delta),
    \\
    \beta_{-}&=\frac{1}{\delta} \beta_{-}^{(-1)}+\beta_{-}^{(0)}+\mathcal{O}(\delta). 
\end{align}
To leading order, we find
\begin{align}
     \label{eq:beta+ large r}
    \beta_{+}&=\sqrt{\frac{\mu_{+}}{\mu_{-}}\chi^{a}\chi_{a}}=\sqrt{\mu_{+}}\sqrt{\chi^{a}\chi_{a}}+\mathcal{O}(\mu_{+}),
    \\
    \label{eq:beta- large r}
    \beta_{-}&=\sqrt{\frac{\mu_{-}}{\mu_{+}}\chi^{a}\chi_{a}}
    =\frac{1}{\sqrt{\mu_{+}}}\sqrt{\chi^{a}\chi_{a}}+\mathcal{O}(1).
\end{align}
It can be seen that $\beta_{\pm}$ do not depend on $\Gamma$ to leading order.

We have now found expressions for $\beta_\pm$ near the light cylinder (\ref{eq:beta- near LC}) and at large $r$ (\ref{eq:beta- large r}).  The transition between these behaviors occurs roughly where the two are equal, 
\begin{align}
    \label{eq:location SA old}
    2\Gamma& =\frac{1}{\sqrt{\mu_{+}}}\sqrt{\chi^{a}\chi_{a}},
\end{align}
which implies
\begin{align}
    4\Gamma^2 \mu_+ = -1+ \frac{2M}{r}+\Omega^2 r^2 \sin^2\! \theta.
\end{align}
For slowly rotating stars such that gravity is negligible at the light cylinder, we may drop the middle term on the RHS and solve as
\begin{align}\label{eq:location SA}
    r = \frac{\sqrt{1+4 \Gamma^2 \mu_+}}{\Omega \sin \theta}.
\end{align}
Since $\mu_+$ is small, we confirm that the transition occurs near the light cylinder $r_{\rm LC}=1/(\Omega \sin \theta)$.

The maximum electric field at this transition radius is roughly set by the electric field value as determined by the large-$r$ profile.  Using Eq.~(\ref{eq: parallel E 1st}) and (\ref{eq:beta- large r}), the parallel electric field in the large $r$ region is given by
\begin{align}
    \label{eq:E para schd mono}
    \delta E_{0}&=-2\ell^{a}D_{a}\left(u_{-}^{a}\chi_{a}\right)\notag
    \\
    &=\partial_{r}\left(-\frac{1}{2}\beta_{-}+\frac{1}{2}n^{a}\chi_{a}\right)\notag
    \\
    &=-\frac{1}{2\sqrt{\mu_{+}}}\partial_{r} \left(\sqrt{\chi^{a}\chi_{a}}\right)+\mathcal{O}(\mu_{+}^{0})\notag
    \\
    &=-\frac{1}{2\sqrt{\mu_{+}}\sqrt{\chi^{a}\chi_{a}}}\left(
    \frac{M}{r^{2}}+r\Omega^{2}\sin^{2}{\theta}
    \right).
\end{align}
The maximum value of $\delta E_{0}$ is obtained by evaluating this at the location of the strong acceleration.  When gravity is negligible at the light cylinder, we have the expression \eqref{eq:location SA} for this radius.  However, for estimating the electric field we do not need to consider the small correction due to $\mu_+$, i.e., we can take the transition radius to be the light cylinder radius $r=1/(\Omega \sin \theta)$. 
Then, we find that the maximum value of $\delta E_{0}$ for small $\mu_{+}$ is given by 
\begin{align}
   \delta E_{0}|_{r_{\mathrm{SA}}}\sim - \frac{1}{\mu_{+}}\frac{1}{4\Gamma} \Omega\sin{\theta}.
\end{align}
Since $\delta E_{0}$ is a first-order quantity and the actual parallel electric field is given by $\epsilon \delta E_{0}$ with $\epsilon=m/e$, $\bar{E}_{0}=\epsilon \delta E_{0}$ at the location of the strong acceleration is given by
\begin{align}
    \label{eq: strongest E at LC}
    \bar{E}_{0}|_{r_{\mathrm{SA}}}=-\frac{1}{4\sqrt{2}} \frac{m\Omega\sin{\theta}}{\Gamma e\lambda_{+}|_{R_{*}}\sqrt{\Delta \gamma_{+}|_{R_{*}}}}.
\end{align}

\section{Ingoing/outgoing Kerr-Schild coordinates}\label{app: Kerr-Schild coords}
The Kerr metric in the Boyer-Lindquist coordinates is given by 
\begin{align}
    \label{eq: Kerr metric}
    ds^{2}&=-\frac{\Sigma \Delta}{A}dt^{2}
    +
    \frac{A}{\Sigma}\sin^{2}{\theta}(d\phi-\Omega_{Z}dt)^{2}
    +\Sigma\left(\frac{dr^{2}}{\Delta}+d\theta^{2}\right),
\end{align}
where
\begin{align}
    \Omega_{Z}=\frac{2Mar}{A},
\end{align}
with
\begin{align}
    \Delta &= r^{2}-2Mr+a^{2},
    \\
    \Sigma&=r^{2}+a^{2}\cos^{2}{\theta},
    \\
    A&=(r^{2}+a^{2})^{2}-a^{2}\Delta \sin^{2}{\theta}.
\end{align}
The outer root of $\Delta$ is the event horizon
\begin{align}
    \label{eq: event horizon}
    r_{H}=M+\sqrt{M^{2}-a^{2}}=2M-\frac{a^{2}}{2M}+\mathcal{O}(a^{4}),
\end{align}
which rotates with angular velocity
\begin{align}
    \label{eq: horizon omega}
    \Omega_{H}
    &=
    \frac{a}{a^{2}+r_{H}^{2}}.
\end{align}
The BL coordinates are irregular on both the past and future horizons.   We may instead change to Kerr-Schild coordinates, which can be defined in two ways, where either the past or the future horizon is regular. 
We denote the ingoing/outgoing Kerr-Schild coordinates as ($\tilde{t}_{\rm in/out},\tilde{r}_{\rm in/out},\tilde{\theta}_{\rm in/out},\tilde{\phi}_{\rm in/out}$) whose coordinate transformationn from the BL coordinates is given by \citep{Komissarov2002nov,Crinquand:2020reu}
\begin{align}
    \label{eq:t in/out}
    \tilde{t}_{\rm in/out}&=t\pm M\ln{\frac{\Delta}{M^{2}}}\pm\frac{M^{2}}{\sqrt{M^{2}-a^{2}}}\ln{\left(\frac{r-r_{+}}{r-r_{-}}\right)},
    \\
    \label{eq:phi in/out}
    \tilde{\phi}_{\rm in/out}&=\phi\pm\frac{a}{\sqrt{M^{2}-a^{2}}}\ln{\left(\frac{r-r_{+}}{r-r_{-}}\right)},
\end{align}
along with $\tilde{r}_{\rm in/out}=r,\tilde{\theta}_{\rm in/out}=\theta$. $r_{\pm}$ is the inner/outer horizon radius. Also, the ingoing/outgoing coordinates correspond to $+/-$ in Eqs.(\ref{eq:t in/out}) and (\ref{eq:phi in/out}). 
The future/past horizon is regular for the ingoing/outgoing Kerr-Schild coordinates. 
In the ingoing coordinates, ingoing null geodesic tangent is given by $\partial_{\tilde{t}_{\rm in}}-\partial_{\tilde{r}_{\rm in}}$ while in the outgoing coordinates, $\partial_{\tilde{t}_{\rm out}}+\partial_{\tilde{r}_{\rm out}}$ is a outgoing null geodesic tangent. 
In other words, the ingoing/outgoing light rays are represented by $\tilde{t}_{\rm in/out}\pm \tilde{r}_{\rm in/out} =\mathrm{const.}$ in the ingoing/outgoing coordinates. 
Let $(u^{t},u^{r},u^{\theta},u^{\phi})$ be the components of the four vector of the plasma (timelike vector) in the BL coordinates. Then, in the Kerr-Schild coordinates, 
\begin{align}\label{drdtKS}
    \frac{dr}{d\tilde{t}_{\rm in/out}}=\frac{u^{r}}{u^{t}\pm\frac{2Mr}{\Delta}u^{r}}=\frac{u^{r}}{u^{t}\pm\frac{2M}{r-2M}u^{r}}+\mathcal{O}(a^{2}).
\end{align}
In the main body, we use the ingoing/outgoing Kerr-Schild coordinates for the ingoing/outgoing plasma flows. 
Though this is an arbitrary choice of coordinates, it helps to see the velocities of the plasma relative to the speed of light, since the null rays are given by $dr/d\tilde{t}_{\rm in/out}=\mp1$.

\section{3+1 formulation in flat spacetime}\label{app: define beta}
In this appendix, we introduce a small mass expansion of the two-fluid theory in a flat spacetime using the 3+1 notation. 
This appendix is meant to provide the connection between the 3+1 formulation and the covariant formulation presented in the main part of this paper. 
In this way, readers who are not familiar with the covariant formulation can grasp the main idea behind the method of putting plasma back in the force-free system.  
In the three-velocity formulation, the two-fluid model is described by the following equations:
\begin{align}
    \label{eq: EoM 3d}
    \varepsilon\left\{\frac{\partial(\bar{\gamma}_{\pm}\bar{\vec{v}}_{\pm})}{\partial t} +(\bar{\vec{v}}_{\pm}\cdot\nabla)(\bar{\gamma}_{\pm}\bar{\vec{v}}_{\pm})\right\}&=\pm (\bar{\vec{E}}+\bar{\vec{v}}_{\pm}\times\bar{\vec{B}}),
    \\
    \label{eq: ampare 3d}
    \nabla\times\bar{\vec{B}}-\frac{\partial \bar{\vec{E}}}{\partial t}&=e\bar{n}_{+}\bar{\vec{v}}_{+}-e\bar{n}_{-}\bar{\vec{v}}_{-},
    \\
    \label{eq: Gauss E 3d}
    \nabla \cdot\bar{\vec{E}}&=e\bar{n}_{+}-e\bar{n}_{-},
    \\
    \label{eq: Faraday 3d}
    \nabla\times\bar{\vec{E}}+\frac{\partial \bar{\vec{B}}}{\partial t}&=0,
    \\
    \label{eq: Gauss B 3d}
    \nabla\cdot\bar{\vec{B}}&=0,
    \\
    \label{eq: continuity 3d}
    \frac{\partial \bar{n}_{\pm}}{\partial t}+\nabla\cdot(\bar{n}_{\pm}\bar{\vec{v}}_{\pm})&=0.
\end{align}
where $\bar{\gamma}_{\pm}=1/\sqrt{1-\bar{\vec{v}}^{2}}$. 
Since all variables that appear in the above equations are non-perturbative, we write overbars explicitly to emphasize this point.  In this appendix, $\bar{n}_{\pm}$ are the \textit{lab frame} number densities of positrons and electrons, rather than the rest frame number densities used in the main body.

As in the main body, we consider small $\varepsilon$ and expand all quantities in the power of $\varepsilon$ as
\begin{align}
    \bar{\vec{E}}&=\vec{E}+\varepsilon \delta\vec{E}+\cdots,
    \\
     \bar{\vec{B}}&=\vec{B}+\varepsilon \delta\vec{B}+\cdots,
     \\
      \bar{\vec{v}}_{\pm}&=\vec{v}_{\pm}+\varepsilon \delta\vec{v}_{\pm}+\cdots,
      \\
      \bar{n}_{\pm}&=n_{\pm}+\varepsilon \delta n_{\pm}+\cdots.
\end{align}
To zeroth order, we have the following equations:
\begin{align}
    \label{eq: 0th EoM 3d}
    \vec{E} +\vec{v}_{\pm} \times\vec{B} &=0,
    \\
    \label{eq: 0th ampare 3d}
    \nabla\times\vec{B} -\frac{\partial \vec{E} }{\partial t}&=en _{+}\vec{v} _{+}-en _{-}\vec{v} _{-},
    \\
    \label{eq: 0th Gauss E 3d}
    \nabla\cdot\vec{E} &=en _{+}-en _{-},
    \\
    \label{eq: 0th Faraday E 3d}
    \nabla\times \vec{E} +\frac{\partial\vec{B} }{\partial t}&=0,
    \\
    \label{eq: 0th Gauss B 3d}
    \nabla\cdot\vec{B} &=0,
    \\
    \label{eq: 0th continuity 3d}
    \frac{\partial n_{\pm} }{\partial t}+\nabla\cdot(n_{\pm} \vec{v}_{\pm} )&=0
\end{align}
We can eliminate the plasma degree of freedom ($\vec{v}_{\pm},n_{\pm}$) as follows. By crossing Eq.~(\ref{eq: 0th ampare 3d}) with $\vec{B}$, we find,
\begin{align}
    &\left(\nabla\times\vec{B}-\frac{\partial \vec{E}}{\partial t}\right)\times \vec{B}\notag
    \\
    &=en_{+}\vec{v}_{+}\times\vec{B}-en_{-}\vec{v}_{-}\times\vec{B}\notag
    \\
    &=-en_{+}\vec{E}+en_{-}\vec{E}\notag
    \\
    &=-(en_{+}-en_{-})\vec{E}\notag
    \\
    &=-(\nabla\cdot\vec{E})\vec{E}.
\end{align}
Therefore, the equations that $\vec{E}$ and $\vec{B}$ follow are given by
\begin{align}
    \label{eq: forcefree 3+1 form}
     (\nabla\cdot\vec{E})\cdot\vec{E}+ \left( \nabla\times\vec{B}-\frac{\partial \vec{E}}{\partial t}\right)\times \vec{B}&=0,
     \\
      \nabla\times \vec{E}+\frac{\partial\vec{B}}{\partial t}=0, 
      \quad
    \nabla\cdot\vec{B}&=0.
\end{align}
These equations describe a force-free system since Eq.~(\ref{eq: forcefree 3+1 form}) is nothing but the vanishing condition of the Lorentz force density.

Now, let us consider the first-order EoM for the plasma, which is given by 
\begin{align}
    &\left(\frac{\partial}{\partial t}+\vec{v}_{\pm}\cdot\nabla\right)(\gamma_{\pm}\vec{v}_{\pm})\notag
    \\
    &
    \hspace{1cm}=\pm \left(
    \delta\vec{E}
    +\vec{v}_{\pm} \times\delta\vec{B}
    +\delta \vec{v}_{\pm}\times\vec{B} 
    \right).
\end{align}
By dotting this expression with $\vec{B}$, we have
\begin{align*}
    &\vec{B} \cdot\left[
    \left(\frac{\partial}{\partial t}+\vec{v}_{\pm} \cdot\nabla\right)(\gamma_{\pm} \vec{v}_{\pm} )
    \right]
    \\
    &=\pm \left[\delta\vec{E}\cdot\vec{B} +\vec{B} \cdot(\vec{v}_{\pm} \times\delta\vec{B})
    +\vec{B} \cdot(\delta\vec{v}_{\pm}\times\vec{B} )
    \right]
    \\
    &=\pm\left[\delta\vec{E}\cdot\vec{B} +\delta\vec{B}\cdot(\vec{B} \times\vec{v}_{\pm} )\right]
    \\
    &=\pm \left(
    \delta\vec{E}\cdot\vec{B} 
    +
    \vec{E} \cdot\delta\vec{B}
    \right)
    \\
    &=\pm \delta(\vec{E}\cdot\vec{B}). 
\end{align*}
Since electrons and positrons have to feel the same electric field, the following condition needs to be satisfied:
\begin{align}
    \label{eq:consistency 3}
     \vec{B} \cdot\left[
    \left(\frac{\partial}{\partial t}+\vec{v}_{+} \cdot\nabla\right)(\gamma_{+} \vec{v}_{+} )+\left(\frac{\partial}{\partial t}+\vec{v}_{-} \cdot\nabla\right)(\gamma_{-} \vec{v}_{-} )
    \right]&=0.
\end{align}
This condition is the 3+1 version of Eq.~(\ref{eq: consistency}).
Note that Eq.~(\ref{eq: 0th EoM 3d}) implies that $\vec{v}_{\pm} $ can be written as
\begin{align}
    \label{eq: velocity3}
    \vec{v}_{\pm} &=\frac{\vec{E} \times\vec{B} }{|\vec{B} |^{2}}+v_{\pm\parallel}\frac{\vec{B} }{|\vec{B} |^{2}},
\end{align}
where $v_{\pm\parallel}$ is a component of velocity along the magnetic field line; thus, we can interpret the motion of particles as being stuck to the field lines. 
Since the current and charge density are determined by solving the force-free equations, we have
\begin{align}
    \label{eq:FF current3}
    \vec{J} \equiv& \nabla\times\vec{B} -\frac{\partial \vec{E} }{\partial t}=en_{+} \vec{v}_{+} -en_{-} \vec{v}_{-} ,
    \\
     \label{eq:FF charge3}
    \rho \equiv&\nabla\cdot\vec{E} =en_{-} -en _{-}.
\end{align}
In addition, the plasma has to satisfy the continuity equation
\begin{align}
    \label{eq: continuity 3}
    \frac{\partial n _{\pm}}{\partial t} +\nabla\cdot(n _{\pm}\vec{v}_{\pm} )&=0.
\end{align}
Using Eqs.(\ref{eq:consistency 3})--(\ref{eq: continuity 3}), we can determine the zeroth-order motion of particles ($v_{\pm\parallel}$ and $n _{\pm}$). 

Next, we consider the first-order correction to the field. The first-order equations are given by 
\begin{align}
    \label{eq: 1th EoM 3d}
    &\left(\frac{\partial}{\partial t}+\vec{v}_{\pm} \cdot\nabla\right)(\gamma_{\pm} \vec{v}_{\pm} )\notag
    \\
    &\hspace{1cm}=\pm \left(
    \delta\vec{E}
    +\vec{v}_{\pm} \times\delta\vec{B}
    +\delta\vec{v}_{\pm}\times\vec{B} 
    \right),
    \\
    \label{eq: 1th ampare 3d}
     &\nabla\times\delta\vec{B}-\frac{\partial \delta\vec{E}}{\partial t}\notag
     \\
     &\hspace{1cm}=en_{+} \delta\vec{v}_{+}-en_{-} \delta\vec{v}_{-}+e\delta n_{+}\vec{v}_{+} -e\delta n_{-}\vec{v}_{-} ,
    \\
    \label{eq: 1th Gauss E 3d}
    &\nabla \cdot\delta\vec{E}=e\delta n_{+}-e\delta n_{-},
    \\
    \label{eq: 1th Faraday 3d}
    &\nabla\times\delta \vec{E}+\frac{\partial \delta\vec{B}}{\partial t}=0,
    \\
    \label{eq: 1th Gauss B 3d}
    &\nabla\cdot\delta\vec{B}=0,
    \\
    \label{eq: 1th continuity 3d}
    &\frac{\partial \delta n_{\pm}}{\partial t}+\nabla\cdot(n_{\pm} \delta\vec{v}_{\pm})
    +\nabla\cdot(\delta n_{\pm}\vec{v}_{\pm} )=0.
\end{align}
In fact, it is possible to eliminate $\delta n_{\pm}$ and $\delta \vec{v}_{\pm}$ from the equations for $\delta \vec{E}$ and $\delta \vec{B}$. 
By computing the following quantity, we find,
\begin{align*}
    &(\nabla\cdot\delta\vec{E})\vec{E} +
    \left(  \nabla\times\delta\vec{B}-\frac{\partial \delta\vec{E}}{\partial t}\right)\times \vec{B} 
    \\
    &=
    (e\delta n_{+}-e\delta n_{-})\vec{E} 
    \\
    &+
    \left(
    en_{+} \delta\vec{v}_{+}-en_{-} \delta\vec{v}_{-}+e\delta n_{+}\vec{v}_{+} -e\delta n_{-}\vec{v}_{-} 
    \right)\times\vec{B} 
    \\
    &=en \delta\vec{v}_{+}\times\vec{B} 
    -en \delta\vec{v}_{-}\times\vec{B} 
    \\
    &=en _{+}\left[
     \left(\frac{\partial}{\partial t}+\vec{v}_{+} \cdot\nabla\right)(\gamma_{+} \vec{v}_{+} )
     -\delta \vec{E}-\vec{v}_{+} \times\delta\vec{B}
    \right]
    \\
    &\qquad
    -en _{-}\left[
     -\left(\frac{\partial}{\partial t}+\vec{v}_{+} \cdot\nabla\right)(\gamma_{+} \vec{v}_{+} )
     -\delta\vec{E}-\vec{v}_{+} \times\delta\vec{B}
    \right]
    \\
    &=
    en _{+}
     \left(\frac{\partial}{\partial t}+\vec{v}_{+} \cdot\nabla\right)(\gamma_{+} \vec{v}_{+} )
     +
     en _{-}
     \left(\frac{\partial}{\partial t}+\vec{v}_{-} \cdot\nabla\right)(\gamma_{-} \vec{v}_{-} )\notag
    \\
    &\qquad
     -\rho \delta\vec{E}-\vec{J} \times \delta\vec{B}.
\end{align*}
Thus, the set of equations governing $\delta\vec{E}$ and $\delta\vec{B}$  satisfy are
\begin{align}
    &(\nabla\cdot\delta\vec{E})\vec{E} +
    \left(  \nabla\times\delta\vec{B}-\frac{\partial \delta\vec{E}}{\partial t}\right)\times \vec{B} 
    +\rho \delta\vec{E}+\vec{J} \times \delta\vec{B}\notag
    \\
    &\hspace{1cm}=
    en _{+}
     \left(\frac{\partial}{\partial t}+\vec{v}_{+} \cdot\nabla\right)(\gamma_{+} \vec{v}_{+})
     \\
     &\hspace{2cm}
     +
     en _{-}
     \left(\frac{\partial}{\partial t}+\vec{v}_{-} \cdot\nabla\right)(\gamma_{-} \vec{v}_{-}),
     \\
   &\hspace{1cm} \nabla\times\delta\vec{E}+\frac{\partial \delta\vec{B}}{\partial t}=0,
    \\
    &\hspace{1cm} 
    \nabla\cdot\delta\vec{B}=0.
\end{align}
These are equations only for the first order field $\delta\vec{E}$ and $\delta\vec{B}$ and correspond to Eqs.~(\ref{eq:maxwell1 1st}) and (\ref{eq:maxwell2 1st}).

\section{Faraday plate}
For the rotating magnetospheres considered in the main body, the parallel electric field $\delta E_{0}$ has non-zero values everywhere in the magnetosphere. This is the generic behavior of nearly force-free solutions, but there can be special cases in which $\delta E_{0}$ becomes identically zero. One such example is the so-called Faraday plate, i.e., a rotating disk-shaped conductor in the presence of a uniform magnetic field in the direction perpendicular to the plate. 
The FF EM field for the Fraday plate is given by 
\begin{align}
    \label{eq: F faraday}
    F=qdz\wedge (d\phi-\Omega d t +\Omega d\rho)
\end{align}
using cylindrical coordinates ($t,\rho,\phi,z$) for flat spacetime. 
The current density is $j^{a}=\nabla_{b}F^{ab}=-2q \Omega (\partial_{t}+\partial_{z})$ and the Killing vector field is given by $\chi=\partial_{t}+\Omega\partial_{\phi}$. 
The PNDs for the Faraday plate are also similar to the PNDs of the monopole and they are given by
\begin{align}
    \label{eq:Faraday l+}
    \ell&=\frac{1}{2}\partial_{t}+\frac{1}{2}\partial_{z},
    \\
    \label{eq:Faraday l-}
    n&=
     \frac{1+\Omega^{2}\rho^{2}}{2}\partial_{t}
     +\Omega \partial_{\phi}
     + \frac{-1+\Omega^{2}\rho^{2}}{2}\partial_{z}.
\end{align}
Although the Faraday plate is almost identical to the Michel monopole solution, there is one critical difference: the Faraday plate has an additional symmetry generated by $\partial_{z}$. 
By assuming that the plasma also has this symmetry, the correction to the parallel electric field is found to be
\begin{align}
    \delta E_{0}=\mp \partial_{z}(u^{a}_{\pm}\chi_{a})=0.
\end{align}
In other words, for the Faraday plate case, it is possible to find a two-fluid flow that is consistent with the FF solution and has exactly zero parallel electric field. More generally, zero parallel electric field correction only arises in a high-multiplicity limit (Sec.~\ref{sec:FFMHD limit}).

\bibliography{FFE}

\providecommand{\href}[2]{#2}\begingroup\raggedright\begin{thebibliography}{10}

\bibitem{Uchida1997aug}
T.~Uchida, ``Theory of force-free electromagnetic fields. {{I}}. {{General}} theory,'' \href{http://dx.doi.org/10.1103/PhysRevE.56.2181}{{\em Physical Review E} {\bfseries 56} no.~2, (Aug., 1997) 2181--2197}.

\bibitem{Uchida1997oct}
T.~Uchida, ``Linear perturbations in force-free black hole magnetospheres --- {{II}}. {{Wave}} propagation,'' \href{http://dx.doi.org/10.1093/mnras/291.1.125}{{\em Monthly Notices of the Royal Astronomical Society} {\bfseries 291} no.~1, (Oct., 1997) 125--144}.

\bibitem{Uchida1998jun}
T.~Uchida, ``The force-free magnetosphere around an oblique rotator,'' \href{http://dx.doi.org/10.1046/j.1365-8711.1998.01528.x}{{\em Monthly Notices of the Royal Astronomical Society} {\bfseries 297} no.~1, (June, 1998) 315--322}.

\bibitem{Komissarov2002nov}
S.~S. Komissarov, ``Time-dependent, force-free, degenerate electrodynamics,'' \href{http://dx.doi.org/10.1046/j.1365-8711.2002.05313.x}{{\em Monthly Notices of the Royal Astronomical Society} {\bfseries 336} no.~3, (Nov., 2002) 759--766}.

\bibitem{Gralla.Jacobson2014dec}
S.~E. Gralla and T.~Jacobson, ``Spacetime approach to force-free magnetospheres,'' \href{http://dx.doi.org/10.1093/mnras/stu1690}{{\em Monthly Notices of the Royal Astronomical Society} {\bfseries 445} no.~3, (Dec., 2014) 2500--2534}.

\bibitem{Palenzuela.Bona.ea2011jun}
C.~Palenzuela, C.~Bona, L.~Lehner, and O.~Reula, ``Robustness of the {{Blandford}}--{{Znajek}} mechanism,'' \href{http://dx.doi.org/10.1088/0264-9381/28/13/134007}{{\em Classical and Quantum Gravity} {\bfseries 28} no.~13, (June, 2011) 134007}.

\bibitem{Pfeiffer.MacFadyen2013jul}
H.~P. Pfeiffer and A.~I. MacFadyen, ``Hyperbolicity of {{Force-Free Electrodynamics}},'' July, 2013.

\bibitem{Carrasco.Reula2016apr}
F.~L. Carrasco and O.~A. Reula, ``Covariant hyperbolization of force-free electrodynamics,'' \href{http://dx.doi.org/10.1103/PhysRevD.93.085013}{{\em Physical Review D} {\bfseries 93} no.~8, (Apr., 2016) 085013}.

\bibitem{Goldreich.Julian1969aug}
P.~Goldreich and W.~H. Julian, ``Pulsar {{Electrodynamics}},'' \href{http://dx.doi.org/10.1086/150119}{{\em The Astrophysical Journal} {\bfseries 157} (Aug., 1969) 869}.

\bibitem{Scharlemann1974oct}
E.~T. Scharlemann, ``Aligned rotating magnetospheres. {{II}}. {{Inclusion}} of inertial forces.,'' \href{http://dx.doi.org/10.1086/153150}{{\em The Astrophysical Journal} {\bfseries 193} (Oct., 1974) 217--223}.

\bibitem{Gralla2019may}
S.~E. Gralla, ``Bosonization of strong-field pair plasma,'' \href{http://dx.doi.org/10.1088/1475-7516/2019/05/002}{{\em Journal of Cosmology and Astroparticle Physics} {\bfseries 2019} no.~05, (May, 2019) 002}.

\bibitem{Gralla.Iqbal2019may}
S.~E. Gralla and N.~Iqbal, ``Effective field theory of force-free electrodynamics,'' \href{http://dx.doi.org/10.1103/PhysRevD.99.105004}{{\em Physical Review D} {\bfseries 99} no.~10, (May, 2019) 105004}.

\bibitem{2022ScPP...12...78I}
N.~{Iqbal}, ``{Effective description of non-equilibrium currents in cold magnetized plasma},'' \href{http://dx.doi.org/10.21468/SciPostPhys.12.2.078}{{\em SciPost Physics} {\bfseries 12} no.~2, (Feb., 2022) 078}.

\bibitem{2022ScPP...12...86I}
N.~{Iqbal} and S.~{Ross}, ``{Towards traversable wormholes from force-free plasmas},'' \href{http://dx.doi.org/10.21468/SciPostPhys.12.3.086}{{\em SciPost Physics} {\bfseries 12} no.~3, (Mar., 2022) 086}, \href{http://arxiv.org/abs/2103.01920}{{\ttfamily arXiv:2103.01920 [hep-th]}}.

\bibitem{Gold1968may}
T.~Gold, ``Rotating {{Neutron Stars}} as the {{Origin}} of the {{Pulsating Radio Sources}},'' \href{http://dx.doi.org/10.1038/218731a0}{{\em Nature} {\bfseries 218} no.~5143, (May, 1968) 731--732}.

\bibitem{Blandford.Znajek1977jul}
R.~D. Blandford and R.~L. Znajek, ``Electromagnetic extraction of energy from {{Kerr}} black holes,'' \href{http://dx.doi.org/10.1093/mnras/179.3.433}{{\em Monthly Notices of the Royal Astronomical Society} {\bfseries 179} no.~3, (July, 1977) 433--456}.

\bibitem{Spitkovsky2006aug}
A.~Spitkovsky, ``Time-dependent {{Force-free Pulsar Magnetospheres}}: {{Axisymmetric}} and {{Oblique Rotators}},'' \href{http://dx.doi.org/10.1086/507518}{{\em The Astrophysical Journal} {\bfseries 648} no.~1, (Aug., 2006) L51}.

\bibitem{Kalapotharakos.Contopoulos2009mar}
C.~Kalapotharakos and I.~Contopoulos, ``Three-dimensional numerical simulations of the pulsar magnetosphere: Preliminary results,'' \href{http://dx.doi.org/10.1051/0004-6361:200810281}{{\em Astronomy \& Astrophysics} {\bfseries 496} no.~2, (Mar., 2009) 495--502}.

\bibitem{Petri2012jul}
J.~P{\'e}tri, ``The pulsar force-free magnetosphere linked to its striped wind: Time-dependent pseudo-spectral simulations,'' \href{http://dx.doi.org/10.1111/j.1365-2966.2012.21238.x}{{\em Monthly Notices of the Royal Astronomical Society} {\bfseries 424} no.~1, (July, 2012) 605--619}.

\bibitem{Palenzuela:2010xn}
C.~Palenzuela, T.~Garrett, L.~Lehner, and S.~L. Liebling, ``{Magnetospheres of Black Hole Systems in Force-Free Plasma},'' \href{http://dx.doi.org/10.1103/PhysRevD.82.044045}{{\em Phys. Rev. D} {\bfseries 82} (2010) 044045}, \href{http://arxiv.org/abs/1007.1198}{{\ttfamily arXiv:1007.1198 [gr-qc]}}.

\bibitem{Li2012}
J.~{Li}, A.~{Spitkovsky}, and A.~{Tchekhovskoy}, ``{Resistive Solutions for Pulsar Magnetospheres},'' \href{http://dx.doi.org/10.1088/0004-637X/746/1/60}{{\em \apj} {\bfseries 746} no.~1, (Feb., 2012) 60}, \href{http://arxiv.org/abs/1107.0979}{{\ttfamily arXiv:1107.0979 [astro-ph.HE]}}.

\bibitem{Tchekhovskoy.Spitkovsky.ea2013aug}
A.~Tchekhovskoy, A.~Spitkovsky, and J.~G. Li, ``Time-dependent {{3D}} magnetohydrodynamic pulsar magnetospheres: Oblique rotators.,'' \href{http://dx.doi.org/10.1093/mnrasl/slt076}{{\em Monthly Notices of the Royal Astronomical Society} {\bfseries 435} (Aug., 2013) L1--L5}.

\bibitem{Philippov.Kramer2022}
A.~Philippov and M.~Kramer, ``Pulsar {{Magnetospheres}} and {{Their Radiation}},'' \href{http://dx.doi.org/10.1146/annurev-astro-052920-112338}{{\em Annual Review of Astronomy and Astrophysics} {\bfseries 60} no.~1, (2022) 495--558}.

\bibitem{Petrova2015jan}
S.~A. Petrova, ``Axisymmetric force-free magnetosphere of a pulsar -- {{II}}. {{Transition}} from the self-consistent two-fluid model,'' \href{http://dx.doi.org/10.1093/mnras/stu2270}{{\em Monthly Notices of the Royal Astronomical Society} {\bfseries 446} no.~3, (Jan., 2015) 2243--2250}.

\bibitem{Petrova2017may}
S.~A. Petrova, ``Two-fluid model of the pulsar magnetosphere represented as an axisymmetric force-free dipole,'' \href{http://dx.doi.org/10.1088/1475-7516/2017/05/041}{{\em Journal of Cosmology and Astroparticle Physics} {\bfseries 2017} no.~05, (May, 2017) 041}.

\bibitem{Michel1973mar}
F.~C. Michel, ``Rotating {{Magnetospheres}}: An {{Exact}} 3-{{D Solution}},'' \href{http://dx.doi.org/10.1086/181169}{{\em The Astrophysical Journal} {\bfseries 180} (Mar., 1973) L133}.

\bibitem{ONeil1968}
T.~M. {O'Neil} and J.~H. {Malmberg}, ``{Transition of the Dispersion Roots from Beam-Type to Landau-Type Solutions},'' \href{http://dx.doi.org/10.1063/1.1692190}{{\em Physics of Fluids} {\bfseries 11} no.~8, (Aug., 1968) 1754--1760}.

\bibitem{Wald1984}
R.~M. Wald, \href{http://dx.doi.org/10.7208/chicago/9780226870373.001.0001}{{\em General {{Relativity}}}}.
\newblock University of Chicago Press, 1984.

\bibitem{Kruskal1958mar}
M.~Kruskal, \href{http://dx.doi.org/10.2172/4332926}{``{{THE GYRATION OF A CHARGED PARTICLE}},''} Tech. Rep. NYO-7903; PM-S-33, Princeton Univ., N.J. Project Matterhorn, Mar., 1958.

\bibitem{Northrop1961jul}
T.~G. Northrop, ``The guiding center approximation to charged particle motion,'' \href{http://dx.doi.org/10.1016/0003-4916(61)90167-1}{{\em Annals of Physics} {\bfseries 15} no.~1, (July, 1961) 79--101}.

\bibitem{Vandervoort1960jul}
P.~O. Vandervoort, ``The relativistic motion of a charged particle in an inhomogeneous electromagnetic field,'' \href{http://dx.doi.org/10.1016/0003-4916(60)90004-X}{{\em Annals of Physics} {\bfseries 10} no.~3, (July, 1960) 401--453}.

\bibitem{Berkowitz.Gardner1959}
J.~Berkowitz and C.~S. Gardner, ``On the asymptotic series expansion of the motion of a charged particle in slowly varying fields,'' \href{http://dx.doi.org/10.1002/cpa.3160120307}{{\em Communications on Pure and Applied Mathematics} {\bfseries 12} no.~3, (1959) 501--512}.

\bibitem{Landau.Lifshitz1975jan}
L.~D. Landau and E.~M. Lifshitz, {\em The Classical Theory of Fields}.
\newblock Butterworth-Heinemann, Jan., 1975.

\bibitem{Cai.Gralla.ea2023sepa}
Y.~Cai, S.~E. Gralla, and V.~Paschalidis, ``Dynamics of ultrarelativistic charged particles with strong radiation reaction. {{II}}. {{Entry}} into {{Aristotelian}} equilibrium,'' \href{http://dx.doi.org/10.1103/PhysRevD.108.063019}{{\em Physical Review D} {\bfseries 108} no.~6, (Sept., 2023) 063019}.

\bibitem{Bacchini.Ripperda.ea2020nov}
F.~Bacchini, B.~Ripperda, A.~A. Philippov, and K.~Parfrey, ``A {{Coupled Guiding Center}}--{{Boris Particle Pusher}} for {{Magnetized Plasmas}} in {{Compact-object Magnetospheres}},'' \href{http://dx.doi.org/10.3847/1538-4365/abb604}{{\em The Astrophysical Journal Supplement Series} {\bfseries 251} no.~1, (Nov., 2020) 10}.

\bibitem{Gralla.Lupsasca.ea2017dec}
S.~E. Gralla, A.~Lupsasca, and A.~Philippov, ``Inclined {{Pulsar Magnetospheres}} in {{General Relativity}}: {{Polar Caps}} for the {{Dipole}}, {{Quadrudipole}}, and {{Beyond}},'' \href{http://dx.doi.org/10.3847/1538-4357/aa978d}{{\em The Astrophysical Journal} {\bfseries 851} no.~2, (Dec., 2017) 137}.

\bibitem{Takahashi.Nitta.ea1990nov}
M.~Takahashi, S.~Nitta, Y.~Tatematsu, and A.~Tomimatsu, ``Magnetohydrodynamic {{Flows}} in {{Kerr Geometry}}: {{Energy Extraction}} from {{Black Holes}},'' \href{http://dx.doi.org/10.1086/169331}{{\em The Astrophysical Journal} {\bfseries 363} (Nov., 1990) 206}.

\bibitem{Hirotani:2006gu}
K.~Hirotani, ``{High-energy emission from pulsar magnetospheres},'' \href{http://dx.doi.org/10.1142/S0217732306020846}{{\em Mod. Phys. Lett. A} {\bfseries 21} (2006) 1319--1337}, \href{http://arxiv.org/abs/astro-ph/0606017}{{\ttfamily arXiv:astro-ph/0606017}}.

\bibitem{Hirotani.Pu2016feb}
K.~Hirotani and H.-Y. Pu, ``{{ENERGETIC GAMMA RADIATION FROM RAPIDLY ROTATING BLACK HOLES}},'' \href{http://dx.doi.org/10.3847/0004-637X/818/1/50}{{\em The Astrophysical Journal} {\bfseries 818} no.~1, (Feb., 2016) 50}.

\bibitem{Lovelace.Mehanian.ea1986sep}
R.~V.~E. Lovelace, C.~Mehanian, C.~M. Mobarry, and M.~E. Sulkanen, ``Theory of {{Axisymmetric Magnetohydrodynamic Flows}}: {{Disks}},'' \href{http://dx.doi.org/10.1086/191132}{{\em The Astrophysical Journal Supplement Series} {\bfseries 62} (Sept., 1986) 1}.

\bibitem{Contopoulos1995jun}
J.~Contopoulos, ``Force-free {{Self-similar Magnetically Driven Relativistic Jets}},'' \href{http://dx.doi.org/10.1086/175768}{{\em The Astrophysical Journal} {\bfseries 446} (June, 1995) 67}.

\bibitem{Contopoulos.Kazanas.ea1999jan}
I.~Contopoulos, D.~Kazanas, and C.~Fendt, ``The {{Axisymmetric Pulsar Magnetosphere}},'' \href{http://dx.doi.org/10.1086/306652}{{\em The Astrophysical Journal} {\bfseries 511} no.~1, (Jan., 1999) 351}.

\bibitem{Lyutikov:2011tq}
M.~Lyutikov, ``{Electromagnetic power of merging and collapsing compact objects},'' \href{http://dx.doi.org/10.1103/PhysRevD.83.124035}{{\em Phys. Rev. D} {\bfseries 83} (2011) 124035}, \href{http://arxiv.org/abs/1104.1091}{{\ttfamily arXiv:1104.1091 [astro-ph.HE]}}.

\bibitem{Brennan.Gralla.ea2013sep}
T.~D. Brennan, S.~E. Gralla, and T.~Jacobson, ``Exact solutions to force-free electrodynamics in black hole backgrounds,'' \href{http://dx.doi.org/10.1088/0264-9381/30/19/195012}{{\em Classical and Quantum Gravity} {\bfseries 30} no.~19, (Sept., 2013) 195012}.

\bibitem{Brennan:2013ppa}
T.~D. Brennan and S.~E. Gralla, ``{On the Magnetosphere of an Accelerated Pulsar},'' \href{http://dx.doi.org/10.1103/PhysRevD.89.103013}{{\em Phys. Rev. D} {\bfseries 89} no.~10, (2014) 103013}, \href{http://arxiv.org/abs/1311.0752}{{\ttfamily arXiv:1311.0752 [astro-ph.HE]}}.

\bibitem{Eddington:1924pmh}
A.~S. Eddington, ``{A Comparison of Whitehead's and Einstein's Formul\ae{}},'' \href{http://dx.doi.org/10.1038/113192a0}{{\em Nature} {\bfseries 113} no.~2832, (1924) 192--192}.

\bibitem{Finkelstein1958may}
D.~Finkelstein, ``Past-future asymmetry of the gravitational field of a point particle,'' \href{http://dx.doi.org/10.1103/PhysRev.110.965}{{\em Phys. Rev.} {\bfseries 110} (May, 1958) 965--967}. \url{https://link.aps.org/doi/10.1103/PhysRev.110.965}.

\bibitem{Komissarov2004may}
S.~S. Komissarov, ``Electrodynamics of black hole magnetospheres,'' \href{http://dx.doi.org/10.1111/j.1365-2966.2004.07598.x}{{\em Monthly Notices of the Royal Astronomical Society} {\bfseries 350} no.~2, (May, 2004) 427--448}.

\bibitem{Tanabe:2008wm}
K.~Tanabe and S.~Nagataki, ``{Extended monopole solution of the Blandford-Znajek mechanism: Higher order terms for a Kerr parameter},'' \href{http://dx.doi.org/10.1103/PhysRevD.78.024004}{{\em Phys. Rev. D} {\bfseries 78} (2008) 024004}, \href{http://arxiv.org/abs/0802.0908}{{\ttfamily arXiv:0802.0908 [astro-ph]}}.

\bibitem{Armas.Cai.ea2020apr}
J.~Armas, Y.~Cai, G.~Comp{\`e}re, D.~Garfinkle, and S.~E. Gralla, ``Consistent {{Blandford-Znajek}} expansion,'' \href{http://dx.doi.org/10.1088/1475-7516/2020/04/009}{{\em Journal of Cosmology and Astroparticle Physics} {\bfseries 2020} no.~04, (Apr., 2020) 009}.

\bibitem{Camilloni.Dias.ea2022jul}
F.~Camilloni, O.~J.~C. Dias, G.~Grignani, T.~Harmark, R.~Oliveri, M.~Orselli, A.~Placidi, and J.~E. Santos, ``Blandford-{{Znajek}} monopole expansion revisited: Novel non-analytic contributions to the power emission,'' \href{http://dx.doi.org/10.1088/1475-7516/2022/07/032}{{\em Journal of Cosmology and Astroparticle Physics} {\bfseries 2022} no.~07, (July, 2022) 032}.

\bibitem{Levinson.Cerutti2018aug}
A.~Levinson and B.~Cerutti, ``Particle-in-cell simulations of pair discharges in a starved magnetosphere of a {{Kerr}} black hole,'' \href{http://dx.doi.org/10.1051/0004-6361/201832915}{{\em Astronomy \& Astrophysics} {\bfseries 616} (Aug., 2018) A184}.

\bibitem{Chen.Yuan2020jun}
A.~Y. Chen and Y.~Yuan, ``Physics of {{Pair Producing Gaps}} in {{Black Hole Magnetospheres}}. {{II}}. {{General Relativity}},'' \href{http://dx.doi.org/10.3847/1538-4357/ab8c46}{{\em The Astrophysical Journal} {\bfseries 895} no.~2, (June, 2020) 121}.

\bibitem{Crinquand.Cerutti.ea2020apr}
B.~Crinquand, B.~Cerutti, A.~Philippov, K.~Parfrey, and G.~Dubus, ``Multidimensional {{Simulations}} of {{Ergospheric Pair Discharges}} around {{Black Holes}},'' \href{http://dx.doi.org/10.1103/PhysRevLett.124.145101}{{\em Physical Review Letters} {\bfseries 124} no.~14, (Apr., 2020) 145101}.

\bibitem{Yuan:2025war}
Y.~Yuan, A.~Y. Chen, and M.~Luepker, ``{Physics of Pair-producing Gaps in Black Hole Magnetospheres: Two-dimensional General Relativistic Particle-in-cell Simulations},'' \href{http://dx.doi.org/10.3847/1538-4357/adce79}{{\em Astrophys. J.} {\bfseries 985} no.~2, (2025) 159}, \href{http://arxiv.org/abs/2503.08487}{{\ttfamily arXiv:2503.08487 [astro-ph.HE]}}.

\bibitem{Gralla.Zimmerman2016jun}
S.~E. Gralla and P.~Zimmerman, ``Plasma waves and jets from moving conductors,'' \href{http://dx.doi.org/10.1103/PhysRevD.93.123016}{{\em Physical Review D} {\bfseries 93} no.~12, (June, 2016) 123016}.

\bibitem{Gralla.Lupsasca.ea2016dec}
S.~E. Gralla, A.~Lupsasca, and A.~Philippov, ``{{PULSAR MAGNETOSPHERES}}: {{BEYOND THE FLAT SPACETIME DIPOLE}},'' \href{http://dx.doi.org/10.3847/1538-4357/833/2/258}{{\em The Astrophysical Journal} {\bfseries 833} no.~2, (Dec., 2016) 258}.

\bibitem{Crinquand:2020reu}
B.~Crinquand, B.~Cerutti, G.~Dubus, K.~Parfrey, and A.~Philippov, ``{Synthetic gamma-ray light curves of Kerr black hole magnetospheric activity from particle-in-cell simulations},'' \href{http://dx.doi.org/10.1051/0004-6361/202040158}{{\em Astron. Astrophys.} {\bfseries 650} (2021) A163}, \href{http://arxiv.org/abs/2012.09733}{{\ttfamily arXiv:2012.09733 [astro-ph.HE]}}.

\end{thebibliography}\endgroup
\bibliographystyle{utphys}

\end{document}